\documentclass[aps,prl,preprint,groupedaddress]{revtex4-1}

\usepackage{subfigure}
\usepackage{graphicx}
\usepackage{bm}
\usepackage{amsmath,amssymb}
\usepackage{multirow}
\usepackage{threeparttable}
\usepackage{appendix}
\begin{document}


\title{Reference-frame-independent measurement-device-independent quantum key distribution based on polarization multiplexing}


\author{Hongwei Liu\textsuperscript{1,2}}
\author{Jipeng Wang\textsuperscript{1,2}}
\author{Haiqiang Ma\textsuperscript{2}}
 \email{hqma@bupt.edu.cn}
\author{Shihai Sun\textsuperscript{1}}
\email{shsun@nudt.edu.cn}
\affiliation{1. College of Liberal Arts and Science, National University of Defense Technology, Hunan, Changsha 410073, China\\
2. School of Science and State Key Laboratory of Information Photonics and Optical Communications, Beijing University of Posts and Telecommunications, Beijing 100876, China}


\date{\today}

\begin{abstract}
Measurement-device-independent quantum key distribution (MDI-QKD) is proved to be able to eliminate all potential detector side channel attacks. Combining with the reference frame independent (RFI) scheme, the complexity of practical system can be reduced because of the unnecessary alignment for reference frame. Here, based on polarization multiplexing, we propose a time-bin encoding structure, and experimentally demonstrate the RFI-MDI-QKD protocol. Thanks to this, two of the four Bell states can be distinguished, whereas only one is used to generate the secure key in previous RFI-MDI-QKD experiments. As far as we know, this is the first demonstration for RFI-MDI-QKD protocol with clock rate of 50 MHz and distance of more than hundred kilometers between legitimate parties Alice and Bob. In asymptotic case, we experimentally compare RFI-MDI-QKD protocol with the original MDI-QKD protocol at the transmission distance of 160 km, when the different misalignments of the reference frame are deployed. By considering observables and statistical fluctuations jointly, four-intensity decoy-state RFI-MDI-QKD protocol with biased bases is experimentally achieved at the transmission distance of 100km and 120km. The results show the robustness of our scheme, and the key rate of RFI-MDI-QKD can be improved obviously under a large misalignment of the reference frame.
\end{abstract}


\maketitle

\section{\label{sec:level1}Introduction}
In this highly intelligent age, the privacy of information is vital to the personal life, the management of companies and governments. Recently, researchers turned to physical theory, such as quantum physics, rather than the mathematical complexities to find an unconditional security scheme. Such is the significance of quantum key distribution (QKD) \cite{Bennett1984}, which has been attracted widely attention nowadays. Tremendous theoretic and experimental efforts have been made in this field \cite{COWE,Wang2005Beating,DECOY05,RRDPS,Laing2010Reference,Wang20122,Peng2010Decoy,Yuan2008Gigahertz}. 

However, the actual performance of practical apparatuses should be taken into account in a real QKD system, otherwise the gap between theoretical and practical model will weaken its security \cite{TAG04,ATTACK1,PhysRevA.83.062331,wei2017feasible,PhysRevA.73.022320,1367-2630-12-11-113026,PhysRevLett.107.110501,qi2007time,li2011attacking}. There are three main approaches to close this gap. The first one is the security patch \cite{yuan2010avoiding,da2012real}, but it is not universal for all potential and unnoticed security loopholes. The second one is the device-independent QKD (DI-QKD) \cite{acin2007device,gisin2010proposal,curty2011heralded}, which is still challenging with current technology since a loophole-free Bell test is needed \cite{hensen2015loophole}. The third and the most promising approach is measurement device independent QKD (MDI-QKD) \cite{lo2012measurement,braunstein2012side}. It successfully removes all detection-related security loopholes, which means secure key can be generated even when measurement unit is fully controlled by the adversary Eve. Furthermore, with current technology, MDI-QKD can provide a solution to build  more security long-distance key distribution links or metropolitan networks \cite{yin2016measurement,tang2016measurement}.

The merits of MDI-QKD protocol have attracted extensive attention in recent decades, a series of achievements have been made in both theories \cite{xu2014protocol,curty2014finite,xu2015discrete,CVDV,zhou2016making} and experiments \cite{rubenok2013real,da2013proof,liu2013experimental,comandar2016quantum,tang2016experimental}. Since relative phase and time-bins of pulses can be firmly maintained along the transmission, time-bin encoding is a suitable scheme for fiber based QKD system, whereas the polarization of light is not stable due to the birefringence of fiber. It is noted that most of experiments based on time-bin encoding schemes can only distinguish one Bell state, such as $\left| {{\psi ^ - }} \right\rangle $, which will eventually lead a factor of 3/4 loss in the final key. In addition, an active reference frame alignment is needed to ensure the higher secure key rate. Although additional calibration parts appear feasible, they increase the complexity of the MDI-QKD system, which may lead to extra information leakage through these ancillary processes \cite{Jain2011Device}.

As a promising solution to eliminate the requirements for reference frame calibration, reference-frame-independent (RFI) MDI-QKD protocol is proposed \cite{yin2014reference}. As far as we know, only two experimental verifications were made until now \cite{wang2015phase,wang2017measurement}, whose systems are worked at 1 MHz, and the longest distance between Alice and Bob is 20km. The experimental demonstration with a higher clock rate and longer transmission distance is still missing. Furthermore, although simulations are carried out to compare the performance of RFI-MDI-QKD protocol with the original MDI-QKD protocol under the different misalignments of reference frames \cite{zhang2017practical,zhang2017decoy}, a clearly experimental comparison is also missing.

In this paper, we propose an effective time-bin encoding scheme based on the polarization multiplexing. Combining with the efficient detecting scheme proposed in our previous work \cite{tang2016time}, both bell states $\left| {{\psi ^ \pm }} \right\rangle $ can be distinguished, which means the factor of loss in the final key can be reduced to 1/2. The proof-of-principle experiment based on RFI-MDI-QKD protocol over a symmetrical channel is made to show the feasibility of our scheme. The system clock rate is improved to 50 MHz. In asymptotic case, we compare the performance of RFI-MDI-QKD protocol with the original MDI-QKD protocol at the transmission distance of 160 km. The key rate of an order of magnitude higher is achieved for RFI-MDI-QKD protocol when misalignment of the relative reference frame $\beta$ is controlled at 25 degrees. For real-world applications, we deploy decoy-state RFI-MDI-QKD protocol with biased bases proposed in \cite{zhang2017decoy} for our system. By employing an elegant statistical fluctuation analysis proposed in \cite{wang2017measurement}, the positive secure key rates are achieved for $\beta  = {0^ \circ }$ at the transmission distance of 120km and for $\beta  = {25^ \circ }$ at the transmission distance of 100km. We believe this result can further illustrate the feasibility and the merit of RFI-MDI-QKD protocol under the higher clock rate and longer secure transmission distance, especially at the situation when a large misalignment of reference frame occurred. Eliminating the calibration of primary reference frames of the system will definitely reduce the complexity of the realistic setup, and prevent extra information leakage through the ancillary alignment processes.
\section{\label{sec:level2}Protocol}
In both RFI-MDI-QKD and the original MDI-QKD protocol, Alice and Bob are firstly required a random selection in the several mutually orthogonal bases to prepare their phase randomized weak coherent states, which are \emph{Z} basis states ($\left| {\rm{0}} \right\rangle $, $\left| {\rm{1}} \right\rangle $), \emph{X} basis states ($\left| {\rm{ + }} \right\rangle {\rm{ = }}{{\left( {\left| {\rm{0}} \right\rangle {\rm{ + }}\left| {\rm{1}} \right\rangle } \right)}/ {\sqrt {\rm{2}} }}$, $\left|  -  \right\rangle {\rm{ = }}{{\left( {\left| {\rm{0}} \right\rangle  - \left| {\rm{1}} \right\rangle } \right)} / {\sqrt {\rm{2}} }}$) for the original MDI-QKD protocol, and additional \emph{Y} basis sates ($\left| { + i} \right\rangle {\rm{ = }}{{\left( {\left| {\rm{0}} \right\rangle  + i\left| {\rm{1}} \right\rangle } \right)} /{\sqrt {\rm{2}} }}$, $\left| { - i} \right\rangle {\rm{ = }}{{\left( {\left| {\rm{0}} \right\rangle  - i\left| {\rm{1}} \right\rangle } \right)}/{\sqrt {\rm{2}} }}$) are required in RFI-MDI-QKD protocol. They are then send to an untrusted relay Charlie, who performs a Bell state measurement (BSM) and announces the corresponding measurement results. Charlie's measurement will projects the incoming states into one of two Bell states  $\left| {{\psi ^ + }} \right\rangle  = \left( {\left| {01} \right\rangle  + \left| {10} \right\rangle } \right)/\sqrt 2 $ or $\left| {{\psi ^ - }} \right\rangle  = \left( {\left| {01} \right\rangle  - \left| {10} \right\rangle } \right)/\sqrt 2 $. Alice and Bob keep the data that conform to these instances and discard the rest. After basis sifting and error estimation, they can obtain the total counting rate $Q_{{i_A}{i_B}} ^{{\lambda _A}{\lambda _B}}$ and quantum bit error rate (QBER) $E_{{i_A}{i_B}} ^{{\lambda _A}{\lambda _B}}$, where ${\lambda _{A\left( B \right)}} \in \left\{ {\mu_i ,\nu_i ,o} \right\}$ denotes Alice (Bob) randomly prepare their signal states $\mu_i $, decoy states $\nu_i $ for basis $i_{A(B)} \in \left\{ {Z,X,Y} \right\}$, or vacuum states $o$. It is noted that Alice and Bob do not choose any bases for vacuum states. 

If the deviation of the practical reference from the ideal one ${\beta _{A\left( B \right)}}$ is considered, \emph{Z} basis is assumed well aligned (${Z_A} = {Z_B} = Z$), \emph{X} and \emph{Y} bases can be written as follows \cite{yin2014reference,wang2015phase}:
\begin{equation}
\begin{array}{c}
{X_B}  = \cos \beta {X_A} + \sin \beta {Y_A},\\
{Y_B}  = \cos \beta {Y_A} - \sin \beta {Y_A},\\
\beta  = {{\left| {{\beta _A} - {\beta _B}} \right|} \mathord{\left/
 {\vphantom {{\left| {{\beta _A} - {\beta _B}} \right|} 2}} \right.
 \kern-\nulldelimiterspace} 2}.
\end{array}
\label{1}
\end{equation}

The secure key is extracted from the data when both Alice and Bob encode their bits using signal states ($\mu$) in the \emph{Z} basis. The rest of the data are applied to estimate the parameters used in the secure key rate calculation. The secure key rate is given by \cite{lo2012measurement,wang2017measurement}
\begin{equation}
R \ge {P_{zz}}P_{zz}^{\mu \mu }\left\{ {{\mu ^2}{e^{ - 2\mu }}S_{ZZ}^{11,L}\left[ {1 - {I_E}} \right] - Q_{ZZ}^{\mu \mu }fH\left( {E_{ZZ}^{\mu \mu }} \right)} \right\},
\label{3}
\end{equation}
where $S_{ZZ}^{11,L}$ is a lower bound of the yield of single-photon states in \emph{Z} basis, $P_{zz}$ is the probability that both Alice and Bob send the Z basis state, and $P_{zz}^{\mu \mu }$  is the signal state probability when both the Z basis states are sent from Alice (Bob) respectively. Parameter $f$ is the error correction efficiency, and $H\left( x \right) =  - x{\log _2}\left( x \right) - \left( {1 - x} \right){\log _2}\left( {1 - x} \right)$ is the binary Shannon entropy function.

When sources in both Alice and Bob are assumed perfect, Eve's information ${I_E}$ in Eq.(\ref{3}) can be estimated by ${I_E} = H( {e_{XX}^{11,U}} )$ for the original MDI-QKD protocol, where ${e_{XX}^{11,U}}$ is a upper bound of quantum error rate of single-photon states in \emph{X} basis. As for RFI-MDI-QKD protocol, ${I_E}$ can be bounded by \cite{yin2014reference}
\begin{equation}
\begin{aligned}
{I_E} &= ( {1 - e_{ZZ}^{11,U}} )H\left[ {\left( {1 + u} \right)/2} \right] + e_{ZZ}^{11,U}H\left[ {\left( {1 + v} \right)/2} \right],\\
v &= \sqrt {C/2 - {{( {1 - e_{ZZ}^{11,U}} )}^2}{u^2}} /e_{ZZ}^{11,U},\\
u &= \min [ {C/2/( {1 - e_{ZZ}^{11,U}} ),1} ].
\end{aligned}
\end{equation}
Obviously, ${I_E}$ is a function of upper bound of quantum error rate of single-photon states in \emph{Z} basis $e_{ZZ}^{11,U}$ and the quantity $C$. When there is no Eve and other errors, $C$ always equals to 2. In order to upper bound the Eve's information ${I_E}$, the value of $C$ should be lower bounded, it can be estimated by
\begin{equation}
C  \ge  \sum\limits_{\omega '} {\min \left[ {{{(1 - 2e_{\omega '}^{11,U})}^2},{{(1 - 2e_{\omega '}^{11,L})}^2}} \right]},
\label{5}
\end{equation}
where $\omega '  \in \left\{ {{X_A}{X_B},{X_A}{Y_B},{Y_A}{X_B},{Y_A}{Y_B}} \right\}$ and $e_{\omega '}^{11,U(L)}$ is a upper (lower) bound of the quantum error rate of single-photon states when Alice and Bob choose the $\omega '$ basis simultaneously. Note that $E_{{X_A}{Y_B}}^{\mu \mu }$ and $E_{{Y_A}{X_B}}^{\mu \mu }$ are theoretically symmetrical about 0.5. Thus we assume $E_{\omega'} ^{{\lambda _A}{\lambda _B}} \le 0.5$ for simplicity, if not, Bob can simply flip his bits corresponding to the relevant basis \emph{X}, \emph{Y}. In this scenario, the value $C$ can be simplified by $C \ge {\sum\limits_{\omega '}  {( {1 - 2e_{\omega '} ^{11}} )} ^2}$, where $e_{\omega '}^{11} = \min \left\{ {0.5,e_{\omega '}^{11,U}} \right\}$.

\section{\label{sec:level2}Experimental setup}

\begin{figure*}[hbtp]
\centering
\includegraphics[width=0.9\linewidth]{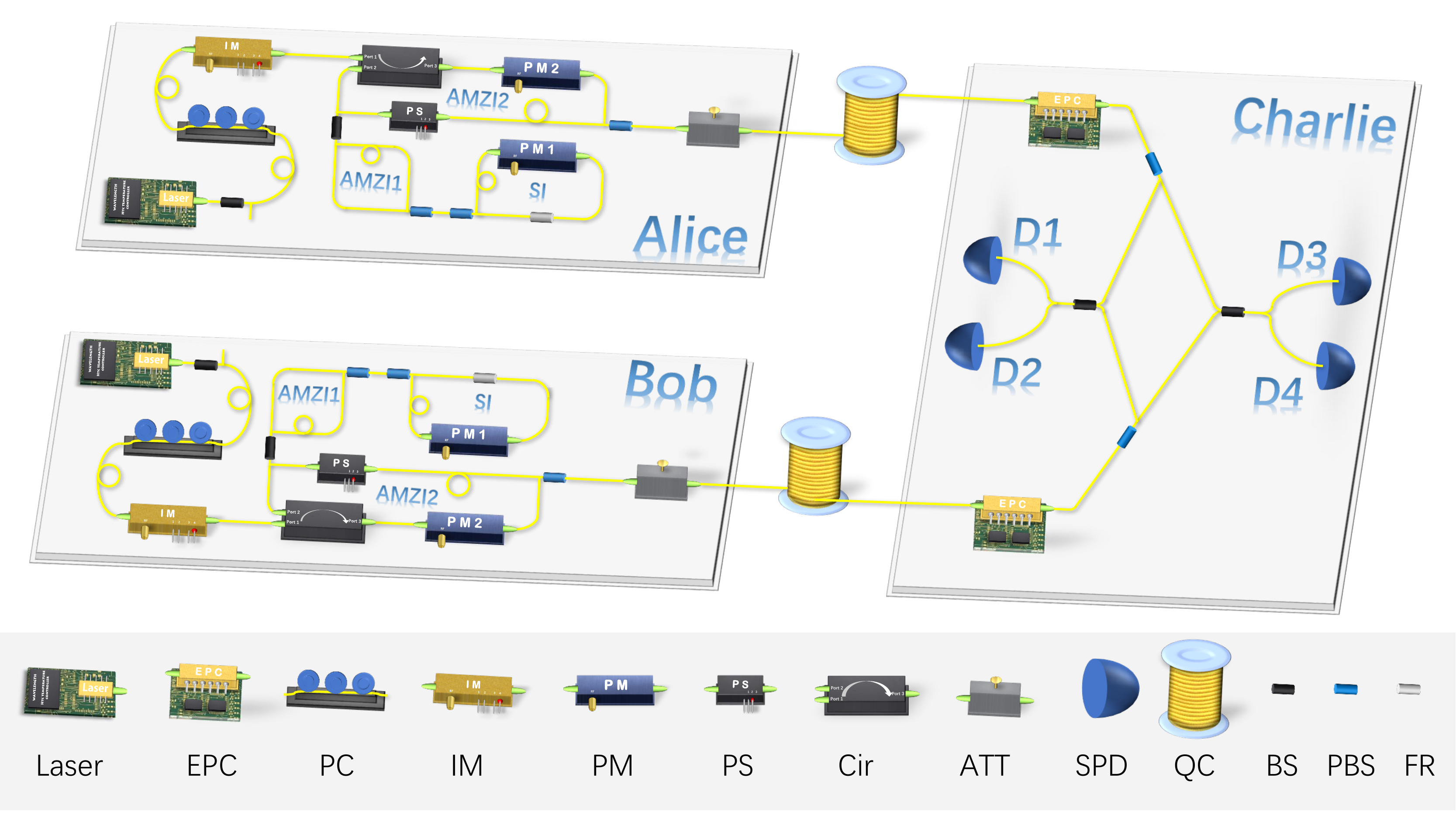}
\caption{Experimental setup of our scheme. Laser, a distributed feedback (DFB) laser combined with a home-built drive board; EPC, electronic polarization controller; PC, polarization controller; IM, intensity modulator; PM, phase modulator; PS, phase shifter; Cir, circulator, its ports and directions is labelled above; ATT, attenuator; SPD, single photon detector; QC, a SMF-28 fiber spool, its channel attenuation is measured at $\alpha=0.195 dB/km$; BS, beam splitter; PBS, polarizing beam splitter; FR, ${90^ \circ }$ Faraday rotator; AMZI, asymmetric Mach-Zehnder interferometer; SI, Sagnac interferometer.}
\label{f1}
\rule{\linewidth}{0.05mm}
\end{figure*}

The time-bin encoding method is used in our system, and the experimental setup is shown in Fig. \ref{f1}. For both Alice and Bob, we employ a distributed feedback (DFB) laser combined with a home-built drive board. By operating the laser below and above the lasering threshold, we first generate phase-randomized laser pulses with a 2 ns temporal width and 50 MHz repetition rate, which eliminates the possibility of an unambiguous-state-discrimination attack \cite{tang2013source}. The electrical pulses are created by an field-programmable gate array (FPGA)-based signal generator (not pictured in Fig. \ref{f1}). In order to calibrate the wavelength, the laser pulses are injected into an optical spectrum analyzer (YOKOGAWA AQ6370D, OSA) through the BSs after two lasers. The OSA, whose resolution is 10-20 pm, is used to monitor the wavelength difference of two independent lasers, which can be minimized by precisely adjusting the operating temperature of lasers through the temperature controllers on the laser drive boards.
\begin{table}[htbp]
\centering
\caption{\bf The detail of time-bin encoding scheme.}
\begin{tabular}{ccccccc}
\hline
 & $\left| {\rm{0}} \right\rangle$ & $\left| {\rm{1}} \right\rangle$  & $\left| {\rm{+}} \right\rangle$ &$\left| {\rm{-}} \right\rangle$  & $\left| {+ i} \right\rangle$ & $\left| {- i} \right\rangle$\\
\hline
PM1 & 0 & $\pi$ & $\frac{\pi }{{\rm{2}}}$ & $\frac{\pi }{{\rm{2}}}$ &  $\frac{\pi }{{\rm{2}}}$ & $\frac{\pi }{{\rm{2}}}$ \\

PM2 &0 &0 &0 & $\pi$ &   $\frac{\pi }{{\rm{2}}}$ & $\frac{3 \pi }{{\rm{2}}}$ \\
\hline
\end{tabular}
\label{t2}
\end{table}

Since Alice's and Bob's parts are symmetrical, here we use Alice's part as an example to illustrate our experimental setups. To realize the decoy states preparation, intensity modulator (IM, Photline, MXER-LN-10) is used to modulate the laser pulses into two different intensities, the vacuum states are prepared by stopping the trigger on lasers. The circulator (Cir) is used to transmit the incident pulses from port 1 to port 2. Each of pulses is then divided into two adjacent pulses with 5 ns separation by first modified asymmetric Mach-Zehnder interferometer (AMZI1), which is composed of a beam splitter (BS) and a polarizing beam splitter (PBS). The relative phase of these two successive pulses is modulated by the phase modulator (PM1, Photline, MPZ-LN-10) in Sagnac interferometer (SI). When the phase of 0 or $\pi$ is modulated, \emph{Z} basis state can be prepared. We define the light passing through the upper path of the second AMZI (AMZI2) as the time-bin state $\left| {\rm{0}} \right\rangle$, and lower path of AMZI2 as the time-bin state $\left| {\rm{1}} \right\rangle$. These two time-bins are separated by 4.2 ns time delay. When the phase of $\pi/2$ is modulated by PM1, the phase modulated by PM2 in AMZI2 are $0$, $\pi$ for \emph{X} basis, and  $\pi/2$, $3\pi/2$ for \emph{Y} basis. For detail, we list our time-bin encoding scheme in Table \ref{t2}.

It is obviously that IM or variable optical attenuator (VOA) is needless to normalize the average photon number of \emph{Z} basis states in two time bins \cite{yin2016measurement,tang2016measurement,wang2017measurement}, these can be achieved only by adjusting the modulating voltage value of PM1 accordingly in our system. This also reduces the complexity of the system to some extent. Furthermore, orthogonal polarization states (\emph{H}, \emph{V}) are multiplexed to the time bins because of the PBS at the output of AMZI2. For the sake of comparing the performance in different misalignment, phase shifters (PS) in AMZI2 are applied to control the reference frame, the parameter of quantum error rate in \emph{X} basis $E_{{X_A}{X_B}}^{{\lambda _A}{\lambda _B}}$ as a guide to set the deviation of relative reference frame. Note that the whole time-bin encoding units are strictly thermal and mechanical isolated to enhance its stability.

At the measurement site, since the time bins are multiplexed with the orthogonal polarization states (\emph{H} and \emph{V}), we can use the PBS to demultiplex them easily. Two electric polarization controllers (EPC, General Photonics, PCD-M02-4X) is used to control the polarization fluctuations that change polarization of input light until the SPD count rate are maximized and hence all polarization changes during photon transmission are compensated for. Two BSs are used to realize the interference. Four commercial InGaAs SPDs (ID210) with an efficiency of ${\eta _d}=12.5\%$, a dark count rate of ${P_d}=1.2 \times {10^{ - 6}}$ and a dead time of 5 $\mu s$ are placed at each output of the BSs. Therefore, all results of BSM are effectively detected, we define the Bell state $\left| {{\psi ^{\rm{ + }}}} \right\rangle $ is D1 and D4 or D2 and D3 in Fig. \ref{f1} clicks simultaneously, and the clicks of D1 and D3 or D2 and D4 is represented by $\left| {{\psi ^ - }} \right\rangle $. The parameters for experiment and numerical simulations are listed in Table \ref{t22}.

\begin{table}[htbp]
\centering
\caption{\bf Parameters for experiment and numerical simulations.}
\begin{tabular}{ccccc}
\hline
 ${\eta _d}$ & ${e_d}$ & $\alpha $  & ${P_d}$ & $f$  \\
\hline
12.5\% & 0.5\% & $0.195dB/km$ & $1.2 \times {10^{ - 6}}$ & 1.16  \\
\hline
\end{tabular}
\label{t22}
\end{table}

\section{Results and discussion}

We first test the indistinguishability of the photons from Alice and Bob by measuring the visibility of Hong-Ou-Mandel (HOM). We obtain a visibility of 42.7\%, which is smaller than the maximally possible value of 50\% for weak coherence source. The low visibility of HOM is mainly caused by detector side imperfections due to after-pulses, it has been studied that after-pulses effect of SPADs has greater impact on the measurement of HOM visibility \cite{Wang:17}. Furthermore, two PBSs are used before interference in our scheme, the change of polarization of incident pulses after long transmission distance will lead to a fluctuating intensity, and the finite extinction ration (about 20dB) of PBS will also lower the visibility. Moreover, the beam splitting ratio and detection efficiency mismatch of detectors can influence the visibility of HOM partly as discussed in \cite{Wang:17}. The central wavelength of laser pulses is 1558.18nm after calibration. Next, we will show and discuss our experimental results for asymptotic case and finite-size pulses case separately.

\subsection{Asymptotic case}
\begin{figure}[ht]
\centering
\includegraphics[width=0.7\linewidth]{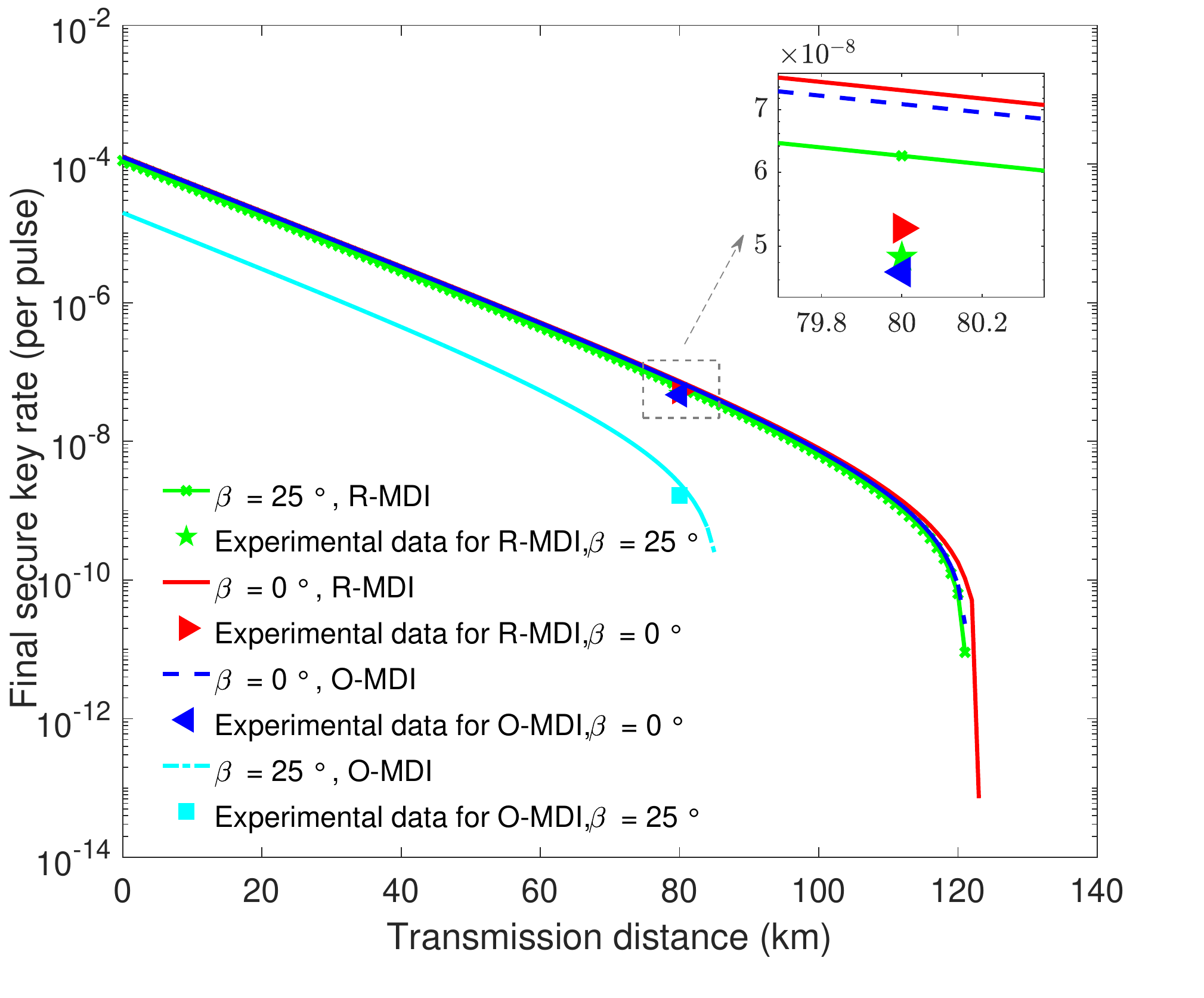}
\caption{Lower secure key rate bound of RFI-MDI-QKD protocol (R-MDI) and the original MDI-QKD protocol (O-MDI) when the deviations of reference frame are controlled at $\beta  ={ 0^ \circ }$ and $\beta  ={ 25^ \circ }$. Except for simulating green curve and purple dashed curve, all average photon number settings of signal state and decoy state are optimized for each of transmission distance. The inserted figure is partial magnification for the experimental results. The horizontal axis represents the distance between Alice (Bob) and Charlie, and quantum channel is symmetric.}
\label{f3a}
\rule{\linewidth}{0.05mm}
\end{figure}

\begin{figure}[htbp]
\centering
\includegraphics[width=0.7\linewidth]{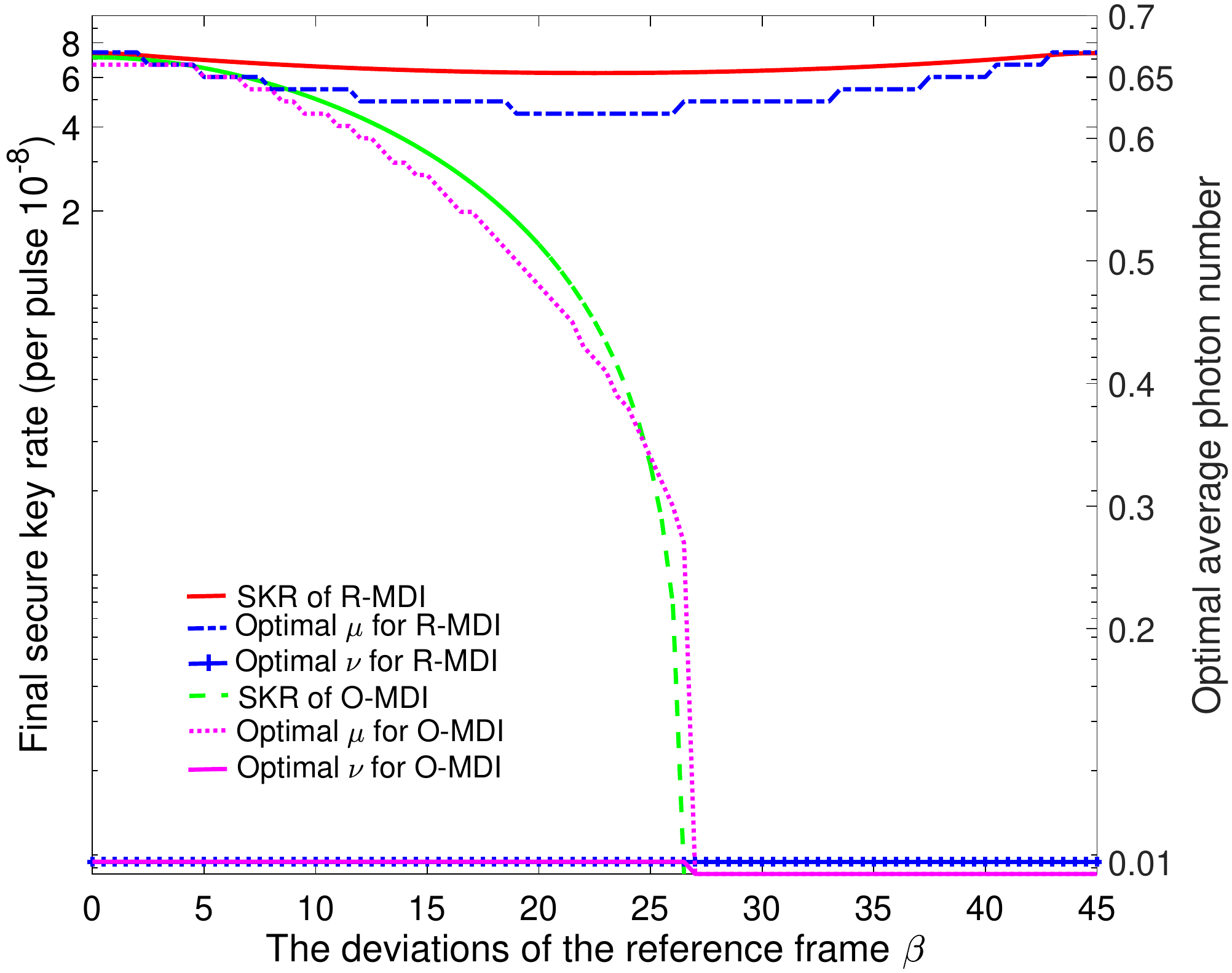}
\caption{Lower secure key rate bound and optimal average photon number of RFI-MDI-QKD protocol (R-MDI) and the original MDI-QKD protocol (O-MDI) versus the different misalignments of reference frame at the distance of 160 km. Since the simulation results are symmetrical about $\beta=45^{\circ}$, this figure only shows the curves at $\beta$ changed from $0^{\circ}$ to $45^{\circ}$.}
\label{f3b}
\rule{\linewidth}{0.05mm}
\end{figure}

In asymptotic case, we adopt symmetrical three-intensity decoy-state protocol for simplicity, which means ${\mu _i} = {\mu _{i'}}{\rm{ = }} \mu $ for signal states and ${\nu _i} = {\nu _{i'}}{\rm{ = }}\nu $ for decoy states. By modeling the total gains and error rates of our system (See Appendices A and B for details), we find the optimal value of average photon number for the original MDI-QKD protocol (O-MDI) and RFI-MDI-QKD protocol (R-MDI) is almost the same, as depicted with blue and purple dash line in Fig. \ref{f3b}, when misalignment of the reference frame is controlled at $\beta  = {0^ \circ }$. This means the secure key rate (SKR) for both protocols can be obtained from a single experiment. The simulation and experimental results are presented in Table \ref{t1} and Fig. \ref{f3a}, which shows two curves are almost overlapped (Red line for R-MDI and blue dash line for O-MDI). We set the average photon number of vacuum state to be $0$ since there is no pulses emitted when the trigger on lasers are paused. The value of $C$ for R-MDI is estimated to 1.668. The QBER of 0.6\% are obtain for \emph{Z} basis, it comes from the successful BSM declared by Charlie when Alice and Bob prepared the same states in Z basis. In the ideal case, the QBER of Z basis should be 0, whereas, the detector's dark counts and finite extinction ratio of the first AMZI in Fig. \ref{f1} will lead to incorrect coincidence counts and thus increase the QBER of Z basis. Meanwhile, the vacuum and multiphoton components of weak coherent states cause accidental coincidences which introduce an error rate of 50\%. Thus, the error rate of the X basis has an expected value of 25\% and so is for Y basis. However, when the visibility of HOM is lower than 50\%, the QBER of X basis will higher than 25\% since the error counts come from the situation when Bell state $\left| {{\psi ^{\rm{ + }}}} \right\rangle $ was announced as Alice and Bob prepared the same states in X basis, or  $\left| {{\psi ^ - }} \right\rangle $ was declared as orthogonal states were prepared. In our system, it is measured at 27.9\%.
\begin{table}[htbp]
\begin{threeparttable}
\centering
\caption{\bf Experimental results when misalignment of reference frame are $\beta  = {0^ \circ }$ and $\beta  = {25^ \circ }$.}
\begin{tabular}{ccccccc}
\hline
Protocol & $\mu $ & $\nu $  & $E_{ZZ}^{\mu \mu }$ & $E_{XX}^{\mu \mu }$  & ${I_E}$ & SKR\\
\hline
\multicolumn{7}{c}{$\beta  = {0^ \circ }$\tnote{*}}\\
\cline{4-5}
R-MDI & \multirow{2}*{0.67} & \multirow{2}*{0.01} & \multirow{2}*{0.006} & \multirow{2}*{0.279} &  0.254 & $5.225 \times {10^{ - 8}}$ \\

O-MDI & & & & &   0.296& $4.690 \times {10^{ - 8}}$ \\
\hline
\multicolumn{7}{c}{$\beta  = {25^ \circ }$}\\
\cline{4-5}
R-MDI & 0.67 & 0.01 & 0.008 & 0.348 &  0.297 & $4.866 \times {10^{ - 8}}$ \\

O-MDI & 0.35 & 0.01 & 0.010 & 0.338 & 0.686   & $1.655\times {10^{ - 9}}$ \\
\hline
\end{tabular}
\begin{tablenotes}
\item[*] The optimal average photon number for O-MDI and R-MDI is identical when misalignment of the reference frame is controlled at $\beta  = {0^ \circ }$. Thus, the SKR for both protocols can be obtained from a single experiment. The $\mu $ and  $\nu $ are optimized in all the test.
\end{tablenotes}
\rule{\linewidth}{0.05mm}
\label{t1}
\end{threeparttable}

\end{table}

In order to investigate the performance of RFI-MDI-QKD protocol and the original MDI-QKD protocol at the nonzero deviation of the relative reference frame, we can control the voltage of PS in Fig. \ref{f1} according to the simulation result of $E_{XX}^{\mu \mu }$ to simulate this deviation. Fig. \ref{f3b} presents the SKR and the optimal average photon number vs different deviation of the reference frame $\beta$ when the transmission distance between Alice and Bob is 160 km. It is obvious that O-MDI is particularly dependent on the change of $\beta$. However, the SKR and the optimal average photon number for R-MDI is almost identical at different deviation of the reference frame. Thus, for R-MDI at $\beta=25^{\circ}$, we keep the values of average photon number in consistency with the setting at $\beta=0^{\circ}$. In this case, the simulation results in Fig. \ref{f3a} show the red curve for $\beta=0^{\circ}$ is almost overlapped with green curve marked with crosses for $\beta=25^{\circ}$. As comparison, the optimal value of $\mu$ and $\nu$ for O-MDI at $\beta=25^{\circ}$ is used to conduct the experimental test. The related experimental results are presented in Table \ref{t1} and Fig. \ref{f3a}. The value of $C$ for R-MDI is estimated to 1.595. At $\beta  ={ 25^ \circ }$, the secure key rate of R-MDI is close to the one at $\beta  ={ 0^ \circ }$, and is an order of magnitude higher than O-MDI at the transmission distance of 160 km. Thus, unlike the O-MDI, the changes of the reference frame nearly cannot influence the secure key rete of R-MDI, neither can optimal average photon number settings. These results well illustrate the robust of RFI-MDI-QKD protocol against the deviation of relative reference frame.

\subsection{Finite-size pulses case}
In real-world applications, the key size is always finite, thus we must consider the effect of statistical fluctuations caused by a finite pulses size. Such an analysis is crucial to ensure the security of RFI-MDI-QKD. Three-intensity decoy-state RFI-MDI-QKD protocol with biased bases proposed in \cite{zhang2017decoy} have been proved that achievable secret key rate and transmission distance can be obviously improved compared with the original protocol, since this protocol avoids the futility in Z basis for decoy states, thus it can simplify the operation of system. Recently, a universal analysis appropriate for fluctuating systems with an arbitrary number of observables is developed in \cite{wang2017measurement}, it is showed that by adopting both the collective constraints and joint parameter estimation techniques, the secret key rate and transmission distance can be impressively improved for four-intensity decoy-state RFI-MDI-QKD protocol. 

\begin{figure}[htbp]
\centering
\includegraphics[width=0.7\linewidth]{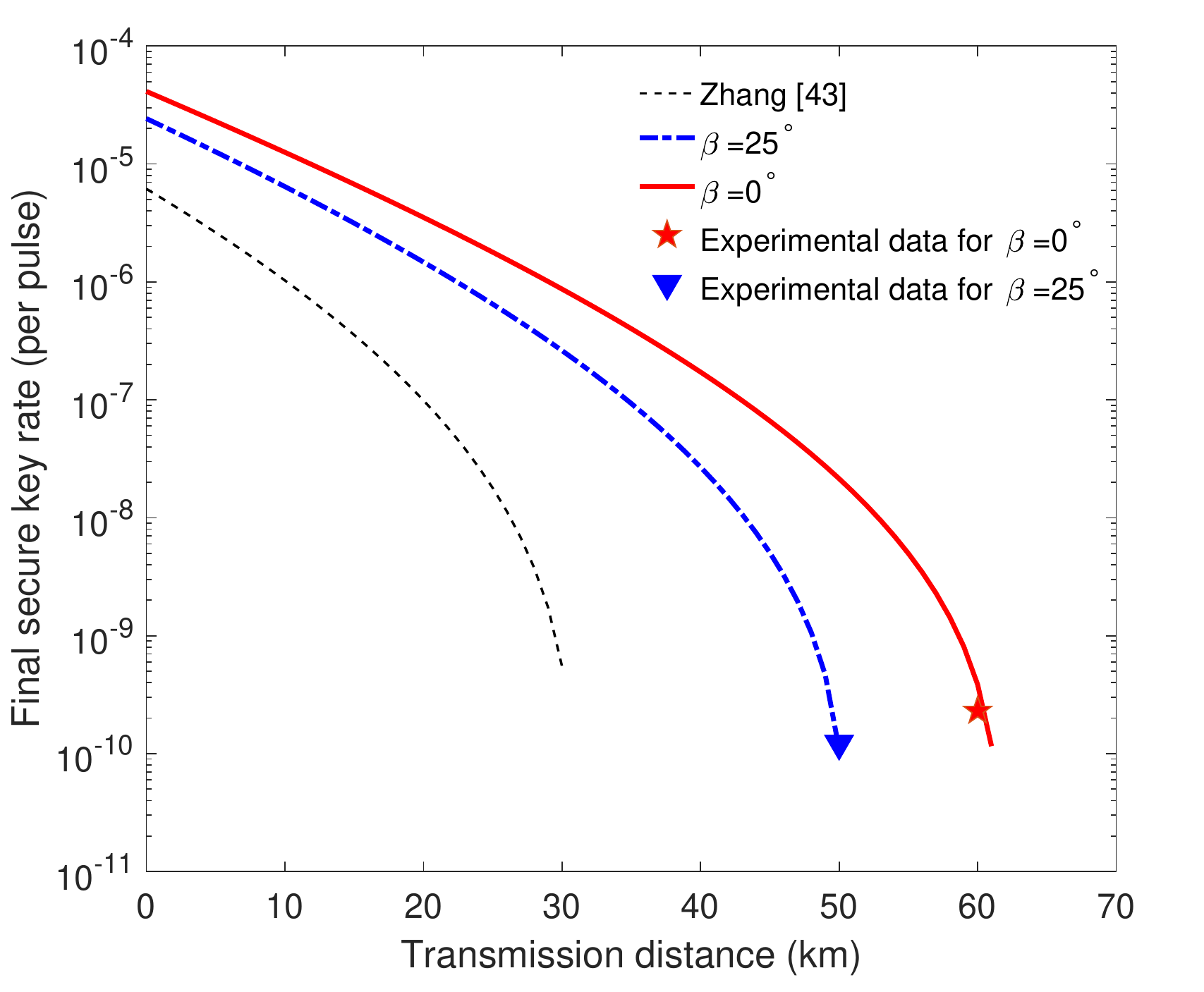}
\caption{Lower secure key rate bound of RFI-MDI-QKD protocol (R-MDI) with biased bases when statistical fluctuations are considered. Black dashed line is the results at $\beta=25^{\circ}$ using the original method proposed in \cite{zhang2017decoy}. The total number of pulse pairs sent from Alice and Bob is $N = 3 \times {10^{12}}$, the failure probability is $\varepsilon  = {10^{ - 10}}$, and the full parameter optimization method is applied. The horizontal axis represents the distance between Alice (Bob) and Charilie, and quantum channel is symmetric.}
\label{f4}
\rule{\linewidth}{0.05mm}
\end{figure}

Here, by using this elegant fluctuation analysis method, we deploy the four-intensity decoy-state RFI-MDI-QKD protocol with biased bases for our experiment. In this scheme, expect for vacuum states, Alice and Bob need to prepare signal states $\mu_z$ for Z basis and $\mu_x$ for both X basis and Y basis due to the symmetry of the X, Y basis in Eq. (\ref{5}), whereas the decoy states $\nu_x$ are prepared only for X basis and Y basis. All related parameter including $\mu_z$, $\mu_x$, $\nu_x$, $P_z$, $P_x$, and $P_{x}^{\mu_x }$ should be optimized to achieve the highest secure key rate. It is found that the achievable secure key rate and transmission distance in this scheme can also be notably improved as showed in Fig. \ref{f4}. 

We apply the Chernoff bound for the fluctuation estimation in our experiment, with a fixed failure probability of $\varepsilon  = {10^{ - 10}}$ and a total number of pulse pairs $N = 3 \times {10^{12}}$. After the simulation with full parameter optimization showed in Fig. \ref{f4}, we find there are some different results compared with the asymptotic case. It is obviously that RFI-MDI-QKD deteriorates with the increase of $\beta$ when statistical fluctuations are considered, which can be explained that the correlations of $e_{{X_A}{X_B}}^{11}$, $e_{{Y_A}{Y_B}}^{11}$, $e_{{X_A}{Y_B}}^{11}$, and $e_{{Y_A}{X_B}}^{11}$ are smeared with the increase of $\beta$, thus it leads to poor estimation of the value of $C$ in Eq. (\ref{5}). Furthermore, the setup of optimal values for experiment will change as the increase of $\beta$, whereas it almost keeps the same in asymptotic case as showed in Fig. \ref{f3b}. For instance, when transmission distance is 100km, the optimal signal intensity setting for Z basis $\mu_z$ at $\beta=0^\circ$ is 0.4407, while it will be 0.2648 if $\beta=25^\circ$. 

\begin{table}[htbp]
\centering
\caption{\bf Experimental results when statistical fluctuations are considered.}
\begin{tabular}{ccccccc}
\hline
 Distance & $\beta$ & ${\mu _{zz}}$  & $E_{ZZ}^{\mu \mu }$ & C &$I_E$  & $SKR$ \\
\hline
100 km & $25^\circ$ & 0.265& 0.9\% & 0.44 & 0.83& $1.22 \times {10^{ - 10}}$ \\

120 km &$0^\circ$ &0.324 &1.15\% & 0.56 & 0.78 & $2.30 \times {10^{ - 10}}$ \\
\hline
\end{tabular}
\label{4}
\end{table}

We experimentally demonstrate the feasible of four-intensity biased decoy-state scheme when statistical fluctuations are considered. Secure key rates for transmission distances of 120 km and 100 km are obtain, which are presented in Table \ref{4} and Fig. \ref{f4}. Their deviations of reference frame are controlled at $\beta=0^\circ$ and $\beta=25^\circ$ respectively.

\section{Conclusion}

In conclusion, a high-speed clock rate of 50 MHz and long distance of more than hundred kilometers RFI-MDI-QKD is demonstrated based on the time-bin and polarization multiplexing. Two of the four Bell states $\left| {{\psi ^ \pm }} \right\rangle $ can be distinguished without a loss. And the states in different bases can be prepared by only using phase modulators without the need for intensity modulators. The value of quantum error rate of Z basis $E_{ZZ}^{\mu \mu }$ shows the feasibility of this scheme. In asymptotic case, we experimentally compare the performance of RFI-MDI-QKD protocol and original MDI-QKD protocol under the difference deviation of reference frame at the distance of 160km. It shows that the secure key rate used RFI-MDI-QKD protocol is an order of magnitude higher than the one used the original MDI-QKD when the misalignment of reference frame is $\beta  = {25^ \circ }$. Moreover, a simulation model for PFI-MDI-QKD protocol is given, and together with the experimental results, the robustness of PFI-MDI-QKD protocol against reference frame change is been verified since the invariant of secure key rate and optimal average photon numbers under the different deviation of the reference frame. The four-intensity decoy-state RFI-MDI-QKD protocol with biased bases is employed to take statistical fluctuations into account in our experiment. By adopting both the collective constraints and joint parameter estimation techniques, the achievable secret key rate and transmission distance is improved obviously compared with the original biased decoy-state RFI-MDI-QKD protocol. We also firstly experimentally achieved this protocol at the transmission distance of 120km when the deviation of reference frame is controlled at $\beta  = {0^ \circ }$ and at the distance of 100km when $\beta  = {25^ \circ }$.

\begin{appendices}
\section*{Appendix A: simulation model}

In order to simulate the protocol performance and get the optimal value of average photon number for experimental system, we need firstly derive the model for total counting rate $Q_{{i_A}{i_B}} ^{{\lambda _A}{\lambda _B}}$ and error counting rate $EQ_{{i_A}{i_B}} ^{{\lambda _A}{\lambda _B}}$. According to the  method in \cite{Ma2012Alternative}, it is deduced by
\begin{equation}
\renewcommand\theequation{A1}
\begin{aligned}
Q_{{Z_A}{Z_B}}^{{\lambda _A}{\lambda _B}} &= {Q_C} + {Q_E},\\
Q_{{Z_A}{Z_B}}^{{\lambda _A}{\lambda _B}}E_{{Z_A}{Z_B}}^{{\lambda _A}{\lambda _B}} & = {e_d}{Q_C} + \left( {1 - {e_d}} \right){Q_E},\\
Q_{{X_A}{X_B}}^{{\lambda _A}{\lambda _B}}  &= 2{y^2}\left[ {2{y^2} - 4y{I_0}\left( x \right) + {I_0}\left( {\rm B} \right) + {I_0}\left( {\rm E} \right)} \right],\\
Q_{{X_A}{X_B}}^{{\lambda _A}{\lambda _B}}E_{{X_A}{X_B}}^{{\lambda _A}{\lambda _B}}  &= 2{y^2}\left[ {{y^2} - 2y{I_0}\left( x \right) + {e_d}{I_0}\left( {\rm B} \right) + \left( {1 - {e_d}} \right){I_0}\left({\rm E} \right)} \right],\\
Q_{{X_A}{Y_B}}^{{\lambda _A}{\lambda _B}}  &= 2{y^2}\left\{ {2{y^2} - 4y{I_0}\left( x \right) + {I_0}\left[  \Theta  \right] + {I_0}\left[ \Xi   \right]} \right\},\\
Q_{{X_A}{Y_B}}^{{\lambda _A}{\lambda _B}}E_{{X_A}{Y_B}}^{{\lambda _A}{\lambda _B}} & = 2{y^2}\left\{ {{y^2} - 2y{I_0}\left( x \right) + {e_d}{I_0}\left[ \Xi  \right] + \left( {1 - {e_d}} \right){I_0}\left[  \Theta  \right]} \right\},\\
Q_{{Y_A}{X_B}}^{{\lambda _A}{\lambda _B}}E_{{Y_A}{X_B}}^{{\lambda _A}{\lambda _B}}  &= 2{y^2}\left\{ {{y^2} - 2y{I_0}\left( x \right) + {e_d}{I_0}\left[ \Theta \right] + \left( {1 - {e_d}} \right){I_0}\left[ \Xi  \right]} \right\},
\end{aligned}
\label{eq:refname1}
\end{equation}
where
\begin{equation}
\renewcommand\theequation{A2}
\begin{array}{l}
{Q_C} = 2{\left( {1 - {P_d}} \right)^2}{e^{{{ - \mu '} \mathord{\left/
 {\vphantom {{ - \mu '} 2}} \right.
 \kern-\nulldelimiterspace} 2}}}[ {1 - \left( {1 - {P_d}} \right){e^{{{ - {\eta _A}{\lambda _A}} \mathord{\left/
 {\vphantom {{ - {\eta _A}{\lambda _A}} 2}} \right.
 \kern-\nulldelimiterspace} 2}}}} ] \\
\times [ {1 - \left( {1 - {P_d}} \right){e^{{{ - {\eta _B}{\lambda _B}} \mathord{\left/
 {\vphantom {{ - {\eta _B}{\lambda _B}} 2}} \right.
 \kern-\nulldelimiterspace} 2}}}} ],\\
{Q_E} = 2{P_d}{\left( {1 - {P_d}} \right)^2}{e^{{{ - \mu '} \mathord{\left/
 {\vphantom {{ - \mu '} 2}} \right.
 \kern-\nulldelimiterspace} 2}}}[ {{I_0}\left( {2x} \right) - \left( {1 - {P_d}} \right){e^{{{ - \mu '} \mathord{\left/
 {\vphantom {{ - \mu '} 2}} \right.
 \kern-\nulldelimiterspace} 2}}}} ],\\
{\rm B} = 2x\cos \beta ,\\
{\rm E} =2x\sin \beta ,\\
\Theta  = \sqrt 2 x\left( {\cos \beta  + \sin \beta } \right),\\
\Xi  = \sqrt 2 x\left( {\cos \beta  - \sin \beta } \right),
\end{array}
\label{eq:refname1}
\end{equation}
${I_0}\left(  \cdot \right)$ is the modified Bessel function of the first kind, ${e_d}$=0.005 is the misalignment-error probability, ${P_d}$ is the dark count of a single-photon detector, ${\eta _A}\left( {{\eta _B}} \right)$ is the transmission of Alice (Bob), and $\mu ' = {\eta _A}{\lambda _A} + {\eta _B}{\lambda _A}$, $x = \sqrt {{\eta _A}{\lambda _A}{\eta _B}{\lambda _B}} /2$ and $y = (1 - {P_d}){e^{-\mu '/4}}$. Due to the symmetry of quantum channel and the \emph{X} ,\emph{Y} basis in Eq. (\ref{5}), we treat the parameters of the \emph{X}, \emph{Y} basis and the average photon number setting for Alice and Bob equivalently for simplicity. Accordingly, ${\mu _A} = {\mu _B}$, $Q_{{X_A}{X_B}}^{{\lambda _A}{\lambda _B}}= Q_{{Y_A}{Y_B}}^{{\lambda _A}{\lambda _B}}$, $Q_{{X_A}{Y_B}}^{{\lambda _A}{\lambda _B}}= Q_{{Y_A}{X_B}}^{{\lambda _A}{\lambda _B}}$, and $EQ_{{X_A}{X_B}}^{{\lambda _A}{\lambda _B}}= EQ_{{Y_A}{Y_B}}^{{\lambda _A}{\lambda _B}}$. The quantum error rate can be calculated by $E_{{i_A}{i_B}} ^{{\lambda _A}{\lambda _b}} = EQ_{{i_A}{i_B}} ^{{\lambda _A}{\lambda _b}}/Q_{{i_A}{i_B}} ^{{\lambda _A}{\lambda _b}}$, it is obvious that $E_{{X_A}{Y_B}}^{{\lambda _A}{\lambda _b}}$ and $E_{{Y_A}{X_B}}^{{\lambda _A}{\lambda _b}}$ is symmetrical about $0.5$. As mentioned above, we assume the quantum error rate is smaller than 0.5 for simplicity, thus $e_{{X_A}{Y_B}}^{{\lambda _A}{\lambda _b}} = 1 - E_{{X_A}{Y_B}}^{{\lambda _A}{\lambda _b}}$ if $E_{{X_A}{Y_B}}^{{\lambda _A}{\lambda _b}} > 0.5$.

\section*{Appendix B: secure key rate estimation}

The secure key rate of Eq. (\ref{3}) is calculated with an analytical method with two decoy states according to \cite{wang2017measurement,yu2013three}.The lower bound and upper bound of the single-photon yield and the error yield is given by
\begin{equation}
\renewcommand\theequation{B1}
\begin{aligned}
{m^{11L}} &\ge \frac{{{T_1} - {T_2} - {a'_1}{b'_2}{T_3}}}{{{a_1}{a'_1}\left( {{b_1}{b'_2} - {b'_1}{b_2}} \right)}},\\
{m^{11U}} &\le \frac{{{M^{{v_i}{v_i}}} - {T_3}}}{{{a_1}{b_1}}},
\label{8} 
\end{aligned}
\end{equation}
where
\begin{equation}
\renewcommand\theequation{B2}
\begin{array}{l}
{T_1} = {a'_1}{b'_2}{M^{{v_i}{v_i}}} + {a_1}{b_2}{a'_0}{M^{o\mu_i }} + {a_1}{b_2}{a'_0}{M^{\mu_i o}},\\
{T_2} = {a_1}{b_2}{M^{{\mu_i}{ \mu_i} }} + {a_1}{b_2}{a'_0}{b'_0}{M^{oo}},\\
{T_3} = {a_0}{M^{o{v_i}}} + {b_0}{M^{{v_i}o}} - {a_0}{b_0}{M^{oo}},\\
a'{\left( {b'} \right)_k} = {{\mu _{i}^k{e^{ - \mu_{i} }}} \mathord{\left/
 {\vphantom {{\mu _{i}^k{e^{ - \mu_{i} }}} {k!}}} \right.
 \kern-\nulldelimiterspace} {k!}},\\
a{\left( b \right)_k} = {{v_{i}^k{e^{ - v_{i}}}} \mathord{\left/
 {\vphantom {{v_{i}^k{e^{ - v_{i}}}} {k!}}} \right.
 \kern-\nulldelimiterspace} {k!}}.
\end{array}
\end{equation}
In the above formula, ${{M^{{\lambda _A}{\lambda _B}}} \in \left\{ {{Q^{{\lambda _A}{\lambda _B}}},E{Q^{{\lambda _A}{\lambda _B}}}} \right\}}$, ${m^{11}} \in \left\{ {{S^{11}},e{S^{11}}} \right\}$, $i \in \left\{ {Z,X,Y} \right\}$, and ${e^{11L(U)}} = e{S^{11L(U)}}/{S^{11U(L)}}$. 

It is noted that the expression of Eq. (B1) is independent on $\omega $, thus, we can use above equations to estimate the parameters in Eq. (\ref{3}) for asymptotic case, which are listed in Table \ref{t5}, and then to calculate the secure key rate. However, since there only is signal states for Z basis in biased decoy-state protocol, we emphasize that $e_{ZZ}^{11U}$ and $e_{XX}^{11U}$ may be different and should be estimated individually. By using the following formula 
\begin{equation}
\renewcommand\theequation{B3}
\begin{aligned}
{m^{11U}} &\le \frac{{{M^{u_{z}u_{z}}} - {T'_3}}}{{{a'_1}{b'_1}}},
\label{B3} 
\end{aligned}
\end{equation}
the upper bound of error yield for Z basis can be estimated. Where
\begin{equation}
\renewcommand\theequation{B4}
\begin{aligned}
{T'_3} = {a'_0}{M^{o{u_{z}}}} + {b'_0}{M^{u_{z}o}} - {a'_0}{b'_0}{M^{oo}}.
\label{B4} 
\end{aligned}
\end{equation}
By using the fluctuation analysis method proposed in \cite{wang2017measurement}, the parameters used for secure key rate estimation are listed in Table \ref{t5}.

\begin{table}[htbp]
\centering
\caption{\bf Parameters estimated in the process of secure key rate calculation.}
\begin{tabular}{ccccccc}
\hline
 $\beta $  & $e_{ZZ}^{11U}$  & $e_{XX}^{11U}$ & $e_{YY}^{11U}$ & $e_{XY}^{11U}$ & $e_{YX}^{11U}$ & $S_{ZZ}^{11L} (10^{ - 6})$  \\
\hline
\multicolumn{7}{c}{R-MDI in asymptotic case}\\
\hline
${0^ \circ }$ & 0.004 & 0.052 & 0.035 &  0.534 & 0.527 & 1.084 \\

 ${25^ \circ }$ & 0.005 & 0.174 & 0.225 &  0.176 & 0.166 & 1.221 \\
\hline
\multicolumn{7}{c}{O-MDI in in asymptotic case }\\
\hline
${0^ \circ }$ & 0.004 & 0.052 & - & - & -& 1.084 \\

 ${25^ \circ }$ & 0.005 & 0.182 &- &  - & - & 1.200 \\
\hline
\multicolumn{7}{c}{R-MDI with biased bases in finite-data case }\\
\hline
${0^ \circ }$ & 0.020 & 0.262 & 0.212 & 0.683 &0.631& 6.959 \\

 ${25^ \circ }$ & 0.015 & 0.348 &0.350 &  0.319 &0.316 & 17.305 \\
\hline
\end{tabular}
\label{t5}
\end{table}
\end{appendices}

\section*{Funding Information}
National Natural Science Foundation of China (NSFC) (11674397); Fund of State Key Laboratory of Information Photonics and Optical Communications (Beijing University of Posts and Telecommunications) (No. IPOC2017ZT04), P. R. China.

\section*{Acknowledgments}
The authors would like to thank Chao Wang for helpful discussion in statistical fluctuations analysis.

\bibliography{sample}

\begin{thebibliography}{49}%
\makeatletter
\providecommand \@ifxundefined [1]{%
 \@ifx{#1\undefined}
}%
\providecommand \@ifnum [1]{%
 \ifnum #1\expandafter \@firstoftwo
 \else \expandafter \@secondoftwo
 \fi
}%
\providecommand \@ifx [1]{%
 \ifx #1\expandafter \@firstoftwo
 \else \expandafter \@secondoftwo
 \fi
}%
\providecommand \natexlab [1]{#1}%
\providecommand \enquote  [1]{``#1''}%
\providecommand \bibnamefont  [1]{#1}%
\providecommand \bibfnamefont [1]{#1}%
\providecommand \citenamefont [1]{#1}%
\providecommand \href@noop [0]{\@secondoftwo}%
\providecommand \href [0]{\begingroup \@sanitize@url \@href}%
\providecommand \@href[1]{\@@startlink{#1}\@@href}%
\providecommand \@@href[1]{\endgroup#1\@@endlink}%
\providecommand \@sanitize@url [0]{\catcode `\\12\catcode `\$12\catcode
  `\&12\catcode `\#12\catcode `\^12\catcode `\_12\catcode `\%12\relax}%
\providecommand \@@startlink[1]{}%
\providecommand \@@endlink[0]{}%
\providecommand \url  [0]{\begingroup\@sanitize@url \@url }%
\providecommand \@url [1]{\endgroup\@href {#1}{\urlprefix }}%
\providecommand \urlprefix  [0]{URL }%
\providecommand \Eprint [0]{\href }%
\providecommand \doibase [0]{http://dx.doi.org/}%
\providecommand \selectlanguage [0]{\@gobble}%
\providecommand \bibinfo  [0]{\@secondoftwo}%
\providecommand \bibfield  [0]{\@secondoftwo}%
\providecommand \translation [1]{[#1]}%
\providecommand \BibitemOpen [0]{}%
\providecommand \bibitemStop [0]{}%
\providecommand \bibitemNoStop [0]{.\EOS\space}%
\providecommand \EOS [0]{\spacefactor3000\relax}%
\providecommand \BibitemShut  [1]{\csname bibitem#1\endcsname}%
\let\auto@bib@innerbib\@empty
\bibitem [{\citenamefont {Bennett}\ and\ \citenamefont
  {Brassard}(1984)}]{Bennett1984}%
  \BibitemOpen
  \bibfield  {author} {\bibinfo {author} {\bibfnamefont {C.~H.}\ \bibnamefont
  {Bennett}}\ and\ \bibinfo {author} {\bibfnamefont {G.}~\bibnamefont
  {Brassard}},\ }in\ \href {\doibase 10.1016/j.tcs.2011.08.039} {\emph
  {\bibinfo {booktitle} {Proc. 1984 IEEE International Conference on Computers,
  Systems, and Signal Processing}}}\ (\bibinfo {year} {1984})\ pp.\ \bibinfo
  {pages} {175--179}\BibitemShut {NoStop}%
\bibitem [{\citenamefont {Stucki}\ \emph {et~al.}(2009)\citenamefont {Stucki},
  \citenamefont {Walenta}, \citenamefont {Vannel}, \citenamefont {Thew},
  \citenamefont {Gisin}, \citenamefont {Zbinden}, \citenamefont {Gray},
  \citenamefont {Towery},\ and\ \citenamefont {Ten}}]{COWE}%
  \BibitemOpen
  \bibfield  {author} {\bibinfo {author} {\bibfnamefont {D.}~\bibnamefont
  {Stucki}}, \bibinfo {author} {\bibfnamefont {N.}~\bibnamefont {Walenta}},
  \bibinfo {author} {\bibfnamefont {F.}~\bibnamefont {Vannel}}, \bibinfo
  {author} {\bibfnamefont {R.~T.}\ \bibnamefont {Thew}}, \bibinfo {author}
  {\bibfnamefont {N.}~\bibnamefont {Gisin}}, \bibinfo {author} {\bibfnamefont
  {H.}~\bibnamefont {Zbinden}}, \bibinfo {author} {\bibfnamefont
  {S.}~\bibnamefont {Gray}}, \bibinfo {author} {\bibfnamefont {C.~R.}\
  \bibnamefont {Towery}}, \ and\ \bibinfo {author} {\bibfnamefont
  {S.}~\bibnamefont {Ten}},\ }\href
  {http://stacks.iop.org/1367-2630/11/i=7/a=075003} {\bibfield  {journal}
  {\bibinfo  {journal} {New Journal of Physics}\ }\textbf {\bibinfo {volume}
  {11}},\ \bibinfo {pages} {075003} (\bibinfo {year} {2009})}\BibitemShut
  {NoStop}%
\bibitem [{\citenamefont {Wang}(2005)}]{Wang2005Beating}%
  \BibitemOpen
  \bibfield  {author} {\bibinfo {author} {\bibfnamefont {X.~B.}\ \bibnamefont
  {Wang}},\ }\href@noop {} {\bibfield  {journal} {\bibinfo  {journal} {Physical
  Review Letters}\ }\textbf {\bibinfo {volume} {94}},\ \bibinfo {pages}
  {230503} (\bibinfo {year} {2005})}\BibitemShut {NoStop}%
\bibitem [{\citenamefont {Lo}\ \emph {et~al.}(2005)\citenamefont {Lo},
  \citenamefont {Ma},\ and\ \citenamefont {Chen}}]{DECOY05}%
  \BibitemOpen
  \bibfield  {author} {\bibinfo {author} {\bibfnamefont {H.-K.}\ \bibnamefont
  {Lo}}, \bibinfo {author} {\bibfnamefont {X.}~\bibnamefont {Ma}}, \ and\
  \bibinfo {author} {\bibfnamefont {K.}~\bibnamefont {Chen}},\ }\href {\doibase
  10.1103/PhysRevLett.94.230504} {\bibfield  {journal} {\bibinfo  {journal}
  {Physical Review Letters}\ }\textbf {\bibinfo {volume} {94}},\ \bibinfo
  {pages} {230504} (\bibinfo {year} {2005})}\BibitemShut {NoStop}%
\bibitem [{\citenamefont {Sasaki}\ \emph {et~al.}(2014)\citenamefont {Sasaki},
  \citenamefont {Yamamoto},\ and\ \citenamefont {Koashi}}]{RRDPS}%
  \BibitemOpen
  \bibfield  {author} {\bibinfo {author} {\bibfnamefont {T.}~\bibnamefont
  {Sasaki}}, \bibinfo {author} {\bibfnamefont {Y.}~\bibnamefont {Yamamoto}}, \
  and\ \bibinfo {author} {\bibfnamefont {M.}~\bibnamefont {Koashi}},\ }\href
  {\doibase 10.1038/nature13303} {\bibfield  {journal} {\bibinfo  {journal}
  {Nature}\ }\textbf {\bibinfo {volume} {509}},\ \bibinfo {pages} {475}
  (\bibinfo {year} {2014})},\ \Eprint {http://arxiv.org/abs/1505.07884}
  {arXiv:1505.07884} \BibitemShut {NoStop}%
\bibitem [{\citenamefont {Laing}\ \emph {et~al.}(2010)\citenamefont {Laing},
  \citenamefont {Scarani}, \citenamefont {Rarity},\ and\ \citenamefont
  {O'Brien}}]{Laing2010Reference}%
  \BibitemOpen
  \bibfield  {author} {\bibinfo {author} {\bibfnamefont {A.}~\bibnamefont
  {Laing}}, \bibinfo {author} {\bibfnamefont {V.}~\bibnamefont {Scarani}},
  \bibinfo {author} {\bibfnamefont {J.~G.}\ \bibnamefont {Rarity}}, \ and\
  \bibinfo {author} {\bibfnamefont {J.~L.}\ \bibnamefont {O'Brien}},\
  }\href@noop {} {\bibfield  {journal} {\bibinfo  {journal} {Physical Review
  A}\ }\textbf {\bibinfo {volume} {82}},\ \bibinfo {pages} {7261} (\bibinfo
  {year} {2010})}\BibitemShut {NoStop}%
\bibitem [{\citenamefont {Wang}\ \emph {et~al.}(2012)\citenamefont {Wang},
  \citenamefont {Chen}, \citenamefont {Guo}, \citenamefont {Yin}, \citenamefont
  {Li}, \citenamefont {Zhou}, \citenamefont {Guo},\ and\ \citenamefont
  {Han}}]{Wang20122}%
  \BibitemOpen
  \bibfield  {author} {\bibinfo {author} {\bibfnamefont {S.}~\bibnamefont
  {Wang}}, \bibinfo {author} {\bibfnamefont {W.}~\bibnamefont {Chen}}, \bibinfo
  {author} {\bibfnamefont {J.~F.}\ \bibnamefont {Guo}}, \bibinfo {author}
  {\bibfnamefont {Z.~Q.}\ \bibnamefont {Yin}}, \bibinfo {author} {\bibfnamefont
  {H.~W.}\ \bibnamefont {Li}}, \bibinfo {author} {\bibfnamefont
  {Z.}~\bibnamefont {Zhou}}, \bibinfo {author} {\bibfnamefont {G.~C.}\
  \bibnamefont {Guo}}, \ and\ \bibinfo {author} {\bibfnamefont {Z.~F.}\
  \bibnamefont {Han}},\ }\href@noop {} {\bibfield  {journal} {\bibinfo
  {journal} {Optics Letters}\ }\textbf {\bibinfo {volume} {37}},\ \bibinfo
  {pages} {1008} (\bibinfo {year} {2012})}\BibitemShut {NoStop}%
\bibitem [{\citenamefont {Peng}\ \emph {et~al.}(2010)\citenamefont {Peng},
  \citenamefont {Liang}, \citenamefont {Wang}, \citenamefont {Pan},
  \citenamefont {Wang}, \citenamefont {Chen}, \citenamefont {Yang},
  \citenamefont {Chen}, \citenamefont {Liu},\ and\ \citenamefont
  {Chen}}]{Peng2010Decoy}%
  \BibitemOpen
  \bibfield  {author} {\bibinfo {author} {\bibfnamefont {C.~Z.}\ \bibnamefont
  {Peng}}, \bibinfo {author} {\bibfnamefont {H.}~\bibnamefont {Liang}},
  \bibinfo {author} {\bibfnamefont {J.}~\bibnamefont {Wang}}, \bibinfo {author}
  {\bibfnamefont {J.~W.}\ \bibnamefont {Pan}}, \bibinfo {author} {\bibfnamefont
  {J.~H.}\ \bibnamefont {Wang}}, \bibinfo {author} {\bibfnamefont
  {K.}~\bibnamefont {Chen}}, \bibinfo {author} {\bibfnamefont {L.}~\bibnamefont
  {Yang}}, \bibinfo {author} {\bibfnamefont {L.~K.}\ \bibnamefont {Chen}},
  \bibinfo {author} {\bibfnamefont {S.~B.}\ \bibnamefont {Liu}}, \ and\
  \bibinfo {author} {\bibfnamefont {T.~Y.}\ \bibnamefont {Chen}},\ }\href@noop
  {} {\bibfield  {journal} {\bibinfo  {journal} {Optics Express}\ }\textbf
  {\bibinfo {volume} {18}},\ \bibinfo {pages} {8587} (\bibinfo {year}
  {2010})}\BibitemShut {NoStop}%
\bibitem [{\citenamefont {Yuan}\ \emph {et~al.}(2008)\citenamefont {Yuan},
  \citenamefont {Dixon}, \citenamefont {Dynes}, \citenamefont {Sharpe},\ and\
  \citenamefont {Shields}}]{Yuan2008Gigahertz}%
  \BibitemOpen
  \bibfield  {author} {\bibinfo {author} {\bibfnamefont {Z.~L.}\ \bibnamefont
  {Yuan}}, \bibinfo {author} {\bibfnamefont {A.~R.}\ \bibnamefont {Dixon}},
  \bibinfo {author} {\bibfnamefont {J.~F.}\ \bibnamefont {Dynes}}, \bibinfo
  {author} {\bibfnamefont {A.~W.}\ \bibnamefont {Sharpe}}, \ and\ \bibinfo
  {author} {\bibfnamefont {A.~J.}\ \bibnamefont {Shields}},\ }\href@noop {}
  {\bibfield  {journal} {\bibinfo  {journal} {Applied Physics Letters}\
  }\textbf {\bibinfo {volume} {92}},\ \bibinfo {pages} {175} (\bibinfo {year}
  {2008})}\BibitemShut {NoStop}%
\bibitem [{\citenamefont {Gottesman}\ \emph {et~al.}(2004)\citenamefont
  {Gottesman}, \citenamefont {Lo}, \citenamefont {L\"{u}tkenhaus},\ and\
  \citenamefont {Preskill}}]{TAG04}%
  \BibitemOpen
  \bibfield  {author} {\bibinfo {author} {\bibfnamefont {D.}~\bibnamefont
  {Gottesman}}, \bibinfo {author} {\bibfnamefont {H.-K.}\ \bibnamefont {Lo}},
  \bibinfo {author} {\bibfnamefont {N.}~\bibnamefont {L\"{u}tkenhaus}}, \ and\
  \bibinfo {author} {\bibfnamefont {J.}~\bibnamefont {Preskill}},\ }\href
  {http://dl.acm.org/citation.cfm?id=2011586.2011587} {\bibfield  {journal}
  {\bibinfo  {journal} {Quantum Info. Comput.}\ }\textbf {\bibinfo {volume}
  {4}},\ \bibinfo {pages} {325} (\bibinfo {year} {2004})}\BibitemShut {NoStop}%
\bibitem [{\citenamefont {Lydersen}\ \emph {et~al.}(2010)\citenamefont
  {Lydersen}, \citenamefont {Wiechers}, \citenamefont {Wittmann}, \citenamefont
  {Elser}, \citenamefont {Skaar},\ and\ \citenamefont {Makarov}}]{ATTACK1}%
  \BibitemOpen
  \bibfield  {author} {\bibinfo {author} {\bibfnamefont {L.}~\bibnamefont
  {Lydersen}}, \bibinfo {author} {\bibfnamefont {C.}~\bibnamefont {Wiechers}},
  \bibinfo {author} {\bibfnamefont {C.}~\bibnamefont {Wittmann}}, \bibinfo
  {author} {\bibfnamefont {D.}~\bibnamefont {Elser}}, \bibinfo {author}
  {\bibfnamefont {J.}~\bibnamefont {Skaar}}, \ and\ \bibinfo {author}
  {\bibfnamefont {V.}~\bibnamefont {Makarov}},\ }\href@noop {} {\bibfield
  {journal} {\bibinfo  {journal} {Nature Photonics}\ }\textbf {\bibinfo
  {volume} {4}},\ \bibinfo {pages} {686} (\bibinfo {year} {2010})}\BibitemShut
  {NoStop}%
\bibitem [{\citenamefont {Sun}\ \emph {et~al.}(2011)\citenamefont {Sun},
  \citenamefont {Jiang},\ and\ \citenamefont {Liang}}]{PhysRevA.83.062331}%
  \BibitemOpen
  \bibfield  {author} {\bibinfo {author} {\bibfnamefont {S.-H.}\ \bibnamefont
  {Sun}}, \bibinfo {author} {\bibfnamefont {M.-S.}\ \bibnamefont {Jiang}}, \
  and\ \bibinfo {author} {\bibfnamefont {L.-M.}\ \bibnamefont {Liang}},\ }\href
  {\doibase 10.1103/PhysRevA.83.062331} {\bibfield  {journal} {\bibinfo
  {journal} {Physical Review A}\ }\textbf {\bibinfo {volume} {83}},\ \bibinfo
  {pages} {062331} (\bibinfo {year} {2011})}\BibitemShut {NoStop}%
\bibitem [{\citenamefont {Wei}\ \emph {et~al.}(2017)\citenamefont {Wei},
  \citenamefont {Liu}, \citenamefont {Ma}, \citenamefont {Yang}, \citenamefont
  {Zhang}, \citenamefont {Sun}, \citenamefont {Xiao},\ and\ \citenamefont
  {Ji}}]{wei2017feasible}%
  \BibitemOpen
  \bibfield  {author} {\bibinfo {author} {\bibfnamefont {K.}~\bibnamefont
  {Wei}}, \bibinfo {author} {\bibfnamefont {H.}~\bibnamefont {Liu}}, \bibinfo
  {author} {\bibfnamefont {H.}~\bibnamefont {Ma}}, \bibinfo {author}
  {\bibfnamefont {X.}~\bibnamefont {Yang}}, \bibinfo {author} {\bibfnamefont
  {Y.}~\bibnamefont {Zhang}}, \bibinfo {author} {\bibfnamefont
  {Y.}~\bibnamefont {Sun}}, \bibinfo {author} {\bibfnamefont {J.}~\bibnamefont
  {Xiao}}, \ and\ \bibinfo {author} {\bibfnamefont {Y.}~\bibnamefont {Ji}},\
  }\href@noop {} {\bibfield  {journal} {\bibinfo  {journal} {Scientific
  Reports}\ }\textbf {\bibinfo {volume} {7}},\ \bibinfo {pages} {449} (\bibinfo
  {year} {2017})}\BibitemShut {NoStop}%
\bibitem [{\citenamefont {Gisin}\ \emph {et~al.}(2006)\citenamefont {Gisin},
  \citenamefont {Fasel}, \citenamefont {Kraus}, \citenamefont {Zbinden},\ and\
  \citenamefont {Ribordy}}]{PhysRevA.73.022320}%
  \BibitemOpen
  \bibfield  {author} {\bibinfo {author} {\bibfnamefont {N.}~\bibnamefont
  {Gisin}}, \bibinfo {author} {\bibfnamefont {S.}~\bibnamefont {Fasel}},
  \bibinfo {author} {\bibfnamefont {B.}~\bibnamefont {Kraus}}, \bibinfo
  {author} {\bibfnamefont {H.}~\bibnamefont {Zbinden}}, \ and\ \bibinfo
  {author} {\bibfnamefont {G.}~\bibnamefont {Ribordy}},\ }\href {\doibase
  10.1103/PhysRevA.73.022320} {\bibfield  {journal} {\bibinfo  {journal}
  {Physical Review A}\ }\textbf {\bibinfo {volume} {73}},\ \bibinfo {pages}
  {022320} (\bibinfo {year} {2006})}\BibitemShut {NoStop}%
\bibitem [{\citenamefont {Xu}\ \emph {et~al.}(2010)\citenamefont {Xu},
  \citenamefont {Qi},\ and\ \citenamefont {Lo}}]{1367-2630-12-11-113026}%
  \BibitemOpen
  \bibfield  {author} {\bibinfo {author} {\bibfnamefont {F.}~\bibnamefont
  {Xu}}, \bibinfo {author} {\bibfnamefont {B.}~\bibnamefont {Qi}}, \ and\
  \bibinfo {author} {\bibfnamefont {H.-K.}\ \bibnamefont {Lo}},\ }\href
  {http://stacks.iop.org/1367-2630/12/i=11/a=113026} {\bibfield  {journal}
  {\bibinfo  {journal} {New Journal of Physics}\ }\textbf {\bibinfo {volume}
  {12}},\ \bibinfo {pages} {113026} (\bibinfo {year} {2010})}\BibitemShut
  {NoStop}%
\bibitem [{\citenamefont {Jain}\ \emph
  {et~al.}(2011{\natexlab{a}})\citenamefont {Jain}, \citenamefont {Wittmann},
  \citenamefont {Lydersen}, \citenamefont {Wiechers}, \citenamefont {Elser},
  \citenamefont {Marquardt}, \citenamefont {Makarov},\ and\ \citenamefont
  {Leuchs}}]{PhysRevLett.107.110501}%
  \BibitemOpen
  \bibfield  {author} {\bibinfo {author} {\bibfnamefont {N.}~\bibnamefont
  {Jain}}, \bibinfo {author} {\bibfnamefont {C.}~\bibnamefont {Wittmann}},
  \bibinfo {author} {\bibfnamefont {L.}~\bibnamefont {Lydersen}}, \bibinfo
  {author} {\bibfnamefont {C.}~\bibnamefont {Wiechers}}, \bibinfo {author}
  {\bibfnamefont {D.}~\bibnamefont {Elser}}, \bibinfo {author} {\bibfnamefont
  {C.}~\bibnamefont {Marquardt}}, \bibinfo {author} {\bibfnamefont
  {V.}~\bibnamefont {Makarov}}, \ and\ \bibinfo {author} {\bibfnamefont
  {G.}~\bibnamefont {Leuchs}},\ }\href {\doibase
  10.1103/PhysRevLett.107.110501} {\bibfield  {journal} {\bibinfo  {journal}
  {Physical Review Letters}\ }\textbf {\bibinfo {volume} {107}},\ \bibinfo
  {pages} {110501} (\bibinfo {year} {2011}{\natexlab{a}})}\BibitemShut
  {NoStop}%
\bibitem [{\citenamefont {Qi}\ \emph {et~al.}(2007)\citenamefont {Qi},
  \citenamefont {Fung}, \citenamefont {Lo},\ and\ \citenamefont
  {Ma}}]{qi2007time}%
  \BibitemOpen
  \bibfield  {author} {\bibinfo {author} {\bibfnamefont {B.}~\bibnamefont
  {Qi}}, \bibinfo {author} {\bibfnamefont {C.-H.~F.}\ \bibnamefont {Fung}},
  \bibinfo {author} {\bibfnamefont {H.-K.}\ \bibnamefont {Lo}}, \ and\ \bibinfo
  {author} {\bibfnamefont {X.}~\bibnamefont {Ma}},\ }\href@noop {} {\bibfield
  {journal} {\bibinfo  {journal} {Quantum Information \& Computation}\ }\textbf
  {\bibinfo {volume} {7}},\ \bibinfo {pages} {73} (\bibinfo {year}
  {2007})}\BibitemShut {NoStop}%
\bibitem [{\citenamefont {Li}\ \emph {et~al.}(2011)\citenamefont {Li},
  \citenamefont {Wang}, \citenamefont {Huang}, \citenamefont {Chen},
  \citenamefont {Yin}, \citenamefont {Li}, \citenamefont {Zhou}, \citenamefont
  {Liu}, \citenamefont {Zhang}, \citenamefont {Guo}, \citenamefont {Bao},\ and\
  \citenamefont {Han}}]{li2011attacking}%
  \BibitemOpen
  \bibfield  {author} {\bibinfo {author} {\bibfnamefont {H.-W.}\ \bibnamefont
  {Li}}, \bibinfo {author} {\bibfnamefont {S.}~\bibnamefont {Wang}}, \bibinfo
  {author} {\bibfnamefont {J.-Z.}\ \bibnamefont {Huang}}, \bibinfo {author}
  {\bibfnamefont {W.}~\bibnamefont {Chen}}, \bibinfo {author} {\bibfnamefont
  {Z.-Q.}\ \bibnamefont {Yin}}, \bibinfo {author} {\bibfnamefont {F.-Y.}\
  \bibnamefont {Li}}, \bibinfo {author} {\bibfnamefont {Z.}~\bibnamefont
  {Zhou}}, \bibinfo {author} {\bibfnamefont {D.}~\bibnamefont {Liu}}, \bibinfo
  {author} {\bibfnamefont {Y.}~\bibnamefont {Zhang}}, \bibinfo {author}
  {\bibfnamefont {G.-C.}\ \bibnamefont {Guo}}, \bibinfo {author} {\bibfnamefont
  {W.-S.}\ \bibnamefont {Bao}}, \ and\ \bibinfo {author} {\bibfnamefont
  {Z.-F.}\ \bibnamefont {Han}},\ }\href {\doibase 10.1103/PhysRevA.84.062308}
  {\bibfield  {journal} {\bibinfo  {journal} {Phys. Rev. A}\ }\textbf {\bibinfo
  {volume} {84}},\ \bibinfo {pages} {062308} (\bibinfo {year}
  {2011})}\BibitemShut {NoStop}%
\bibitem [{\citenamefont {Yuan}\ \emph {et~al.}(2010)\citenamefont {Yuan},
  \citenamefont {Dynes},\ and\ \citenamefont {Shields}}]{yuan2010avoiding}%
  \BibitemOpen
  \bibfield  {author} {\bibinfo {author} {\bibfnamefont {Z.}~\bibnamefont
  {Yuan}}, \bibinfo {author} {\bibfnamefont {J.}~\bibnamefont {Dynes}}, \ and\
  \bibinfo {author} {\bibfnamefont {A.}~\bibnamefont {Shields}},\ }\href@noop
  {} {\bibfield  {journal} {\bibinfo  {journal} {Nature Photonics}\ }\textbf
  {\bibinfo {volume} {4}},\ \bibinfo {pages} {800} (\bibinfo {year}
  {2010})}\BibitemShut {NoStop}%
\bibitem [{\citenamefont {da~Silva}\ \emph {et~al.}(2012)\citenamefont
  {da~Silva}, \citenamefont {Xavier}, \citenamefont {Tempor{\~a}o},\ and\
  \citenamefont {von~der Weid}}]{da2012real}%
  \BibitemOpen
  \bibfield  {author} {\bibinfo {author} {\bibfnamefont {T.~F.}\ \bibnamefont
  {da~Silva}}, \bibinfo {author} {\bibfnamefont {G.~B.}\ \bibnamefont
  {Xavier}}, \bibinfo {author} {\bibfnamefont {G.~P.}\ \bibnamefont
  {Tempor{\~a}o}}, \ and\ \bibinfo {author} {\bibfnamefont {J.~P.}\
  \bibnamefont {von~der Weid}},\ }\href@noop {} {\bibfield  {journal} {\bibinfo
   {journal} {Optics Express}\ }\textbf {\bibinfo {volume} {20}},\ \bibinfo
  {pages} {18911} (\bibinfo {year} {2012})}\BibitemShut {NoStop}%
\bibitem [{\citenamefont {Ac{\'\i}n}\ \emph {et~al.}(2007)\citenamefont
  {Ac{\'\i}n}, \citenamefont {Brunner}, \citenamefont {Gisin}, \citenamefont
  {Massar}, \citenamefont {Pironio},\ and\ \citenamefont
  {Scarani}}]{acin2007device}%
  \BibitemOpen
  \bibfield  {author} {\bibinfo {author} {\bibfnamefont {A.}~\bibnamefont
  {Ac{\'\i}n}}, \bibinfo {author} {\bibfnamefont {N.}~\bibnamefont {Brunner}},
  \bibinfo {author} {\bibfnamefont {N.}~\bibnamefont {Gisin}}, \bibinfo
  {author} {\bibfnamefont {S.}~\bibnamefont {Massar}}, \bibinfo {author}
  {\bibfnamefont {S.}~\bibnamefont {Pironio}}, \ and\ \bibinfo {author}
  {\bibfnamefont {V.}~\bibnamefont {Scarani}},\ }\href@noop {} {\bibfield
  {journal} {\bibinfo  {journal} {Physical Review Letters}\ }\textbf {\bibinfo
  {volume} {98}},\ \bibinfo {pages} {230501} (\bibinfo {year}
  {2007})}\BibitemShut {NoStop}%
\bibitem [{\citenamefont {Gisin}\ \emph {et~al.}(2010)\citenamefont {Gisin},
  \citenamefont {Pironio},\ and\ \citenamefont
  {Sangouard}}]{gisin2010proposal}%
  \BibitemOpen
  \bibfield  {author} {\bibinfo {author} {\bibfnamefont {N.}~\bibnamefont
  {Gisin}}, \bibinfo {author} {\bibfnamefont {S.}~\bibnamefont {Pironio}}, \
  and\ \bibinfo {author} {\bibfnamefont {N.}~\bibnamefont {Sangouard}},\
  }\href@noop {} {\bibfield  {journal} {\bibinfo  {journal} {Physical Review
  Letters}\ }\textbf {\bibinfo {volume} {105}},\ \bibinfo {pages} {070501}
  (\bibinfo {year} {2010})}\BibitemShut {NoStop}%
\bibitem [{\citenamefont {Curty}\ and\ \citenamefont
  {Moroder}(2011)}]{curty2011heralded}%
  \BibitemOpen
  \bibfield  {author} {\bibinfo {author} {\bibfnamefont {M.}~\bibnamefont
  {Curty}}\ and\ \bibinfo {author} {\bibfnamefont {T.}~\bibnamefont
  {Moroder}},\ }\href@noop {} {\bibfield  {journal} {\bibinfo  {journal}
  {Physical Review A}\ }\textbf {\bibinfo {volume} {84}},\ \bibinfo {pages}
  {010304} (\bibinfo {year} {2011})}\BibitemShut {NoStop}%
\bibitem [{\citenamefont {Hensen}\ \emph {et~al.}(2015)\citenamefont {Hensen},
  \citenamefont {Bernien}, \citenamefont {Dr{\'{e}}au}, \citenamefont
  {Reiserer}, \citenamefont {Kalb}, \citenamefont {Blok}, \citenamefont
  {Ruitenberg}, \citenamefont {Vermeulen}, \citenamefont {Schouten},
  \citenamefont {Abell{\'{a}}n}, \citenamefont {Amaya}, \citenamefont
  {Pruneri}, \citenamefont {Mitchell}, \citenamefont {Markham}, \citenamefont
  {Twitchen}, \citenamefont {Elkouss}, \citenamefont {Wehner}, \citenamefont
  {Taminiau},\ and\ \citenamefont {Hanson}}]{hensen2015loophole}%
  \BibitemOpen
  \bibfield  {author} {\bibinfo {author} {\bibfnamefont {B.}~\bibnamefont
  {Hensen}}, \bibinfo {author} {\bibfnamefont {H.}~\bibnamefont {Bernien}},
  \bibinfo {author} {\bibfnamefont {A.~E.}\ \bibnamefont {Dr{\'{e}}au}},
  \bibinfo {author} {\bibfnamefont {A.}~\bibnamefont {Reiserer}}, \bibinfo
  {author} {\bibfnamefont {N.}~\bibnamefont {Kalb}}, \bibinfo {author}
  {\bibfnamefont {M.~S.}\ \bibnamefont {Blok}}, \bibinfo {author}
  {\bibfnamefont {J.}~\bibnamefont {Ruitenberg}}, \bibinfo {author}
  {\bibfnamefont {R.~F.~L.}\ \bibnamefont {Vermeulen}}, \bibinfo {author}
  {\bibfnamefont {R.~N.}\ \bibnamefont {Schouten}}, \bibinfo {author}
  {\bibfnamefont {C.}~\bibnamefont {Abell{\'{a}}n}}, \bibinfo {author}
  {\bibfnamefont {W.}~\bibnamefont {Amaya}}, \bibinfo {author} {\bibfnamefont
  {V.}~\bibnamefont {Pruneri}}, \bibinfo {author} {\bibfnamefont {M.~W.}\
  \bibnamefont {Mitchell}}, \bibinfo {author} {\bibfnamefont {M.}~\bibnamefont
  {Markham}}, \bibinfo {author} {\bibfnamefont {D.~J.}\ \bibnamefont
  {Twitchen}}, \bibinfo {author} {\bibfnamefont {D.}~\bibnamefont {Elkouss}},
  \bibinfo {author} {\bibfnamefont {S.}~\bibnamefont {Wehner}}, \bibinfo
  {author} {\bibfnamefont {T.~H.}\ \bibnamefont {Taminiau}}, \ and\ \bibinfo
  {author} {\bibfnamefont {R.}~\bibnamefont {Hanson}},\ }\href@noop {}
  {\bibfield  {journal} {\bibinfo  {journal} {Nature}\ }\textbf {\bibinfo
  {volume} {526}},\ \bibinfo {pages} {682} (\bibinfo {year}
  {2015})}\BibitemShut {NoStop}%
\bibitem [{\citenamefont {Lo}\ \emph {et~al.}(2012)\citenamefont {Lo},
  \citenamefont {Curty},\ and\ \citenamefont {Qi}}]{lo2012measurement}%
  \BibitemOpen
  \bibfield  {author} {\bibinfo {author} {\bibfnamefont {H.-K.}\ \bibnamefont
  {Lo}}, \bibinfo {author} {\bibfnamefont {M.}~\bibnamefont {Curty}}, \ and\
  \bibinfo {author} {\bibfnamefont {B.}~\bibnamefont {Qi}},\ }\href@noop {}
  {\bibfield  {journal} {\bibinfo  {journal} {Physical Review Letters}\
  }\textbf {\bibinfo {volume} {108}},\ \bibinfo {pages} {130503} (\bibinfo
  {year} {2012})}\BibitemShut {NoStop}%
\bibitem [{\citenamefont {Braunstein}\ and\ \citenamefont
  {Pirandola}(2012)}]{braunstein2012side}%
  \BibitemOpen
  \bibfield  {author} {\bibinfo {author} {\bibfnamefont {S.~L.}\ \bibnamefont
  {Braunstein}}\ and\ \bibinfo {author} {\bibfnamefont {S.}~\bibnamefont
  {Pirandola}},\ }\href@noop {} {\bibfield  {journal} {\bibinfo  {journal}
  {Physical Review Letters}\ }\textbf {\bibinfo {volume} {108}},\ \bibinfo
  {pages} {130502} (\bibinfo {year} {2012})}\BibitemShut {NoStop}%
\bibitem [{\citenamefont {Yin}\ \emph {et~al.}(2016)\citenamefont {Yin},
  \citenamefont {Chen}, \citenamefont {Yu}, \citenamefont {Liu}, \citenamefont
  {You}, \citenamefont {Zhou}, \citenamefont {Chen}, \citenamefont {Mao},
  \citenamefont {Huang}, \citenamefont {Zhang}, \citenamefont {Chen},
  \citenamefont {Li}, \citenamefont {Nolan}, \citenamefont {Zhou},
  \citenamefont {Jiang}, \citenamefont {Wang}, \citenamefont {Zhang},
  \citenamefont {Wang},\ and\ \citenamefont {Pan}}]{yin2016measurement}%
  \BibitemOpen
  \bibfield  {author} {\bibinfo {author} {\bibfnamefont {H.-L.}\ \bibnamefont
  {Yin}}, \bibinfo {author} {\bibfnamefont {T.-Y.}\ \bibnamefont {Chen}},
  \bibinfo {author} {\bibfnamefont {Z.-W.}\ \bibnamefont {Yu}}, \bibinfo
  {author} {\bibfnamefont {H.}~\bibnamefont {Liu}}, \bibinfo {author}
  {\bibfnamefont {L.-X.}\ \bibnamefont {You}}, \bibinfo {author} {\bibfnamefont
  {Y.-H.}\ \bibnamefont {Zhou}}, \bibinfo {author} {\bibfnamefont {S.-J.}\
  \bibnamefont {Chen}}, \bibinfo {author} {\bibfnamefont {Y.}~\bibnamefont
  {Mao}}, \bibinfo {author} {\bibfnamefont {M.-Q.}\ \bibnamefont {Huang}},
  \bibinfo {author} {\bibfnamefont {W.-J.}\ \bibnamefont {Zhang}}, \bibinfo
  {author} {\bibfnamefont {H.}~\bibnamefont {Chen}}, \bibinfo {author}
  {\bibfnamefont {M.~J.}\ \bibnamefont {Li}}, \bibinfo {author} {\bibfnamefont
  {D.}~\bibnamefont {Nolan}}, \bibinfo {author} {\bibfnamefont
  {F.}~\bibnamefont {Zhou}}, \bibinfo {author} {\bibfnamefont {X.}~\bibnamefont
  {Jiang}}, \bibinfo {author} {\bibfnamefont {Z.}~\bibnamefont {Wang}},
  \bibinfo {author} {\bibfnamefont {Q.}~\bibnamefont {Zhang}}, \bibinfo
  {author} {\bibfnamefont {X.-B.}\ \bibnamefont {Wang}}, \ and\ \bibinfo
  {author} {\bibfnamefont {J.-W.}\ \bibnamefont {Pan}},\ }\href {\doibase
  10.1103/PhysRevLett.117.190501} {\bibfield  {journal} {\bibinfo  {journal}
  {Phys. Rev. Lett.}\ }\textbf {\bibinfo {volume} {117}},\ \bibinfo {pages}
  {190501} (\bibinfo {year} {2016})}\BibitemShut {NoStop}%
\bibitem [{\citenamefont {Tang}\ \emph
  {et~al.}(2016{\natexlab{a}})\citenamefont {Tang}, \citenamefont {Yin},
  \citenamefont {Zhao}, \citenamefont {Liu}, \citenamefont {Sun}, \citenamefont
  {Huang}, \citenamefont {Zhang}, \citenamefont {Chen}, \citenamefont {Zhang},
  \citenamefont {You}, \citenamefont {Wang}, \citenamefont {Liu}, \citenamefont
  {Lu}, \citenamefont {Jiang}, \citenamefont {Ma}, \citenamefont {Zhang},
  \citenamefont {Chen},\ and\ \citenamefont {Pan}}]{tang2016measurement}%
  \BibitemOpen
  \bibfield  {author} {\bibinfo {author} {\bibfnamefont {Y.-L.}\ \bibnamefont
  {Tang}}, \bibinfo {author} {\bibfnamefont {H.-L.}\ \bibnamefont {Yin}},
  \bibinfo {author} {\bibfnamefont {Q.}~\bibnamefont {Zhao}}, \bibinfo {author}
  {\bibfnamefont {H.}~\bibnamefont {Liu}}, \bibinfo {author} {\bibfnamefont
  {X.-X.}\ \bibnamefont {Sun}}, \bibinfo {author} {\bibfnamefont {M.-Q.}\
  \bibnamefont {Huang}}, \bibinfo {author} {\bibfnamefont {W.-J.}\ \bibnamefont
  {Zhang}}, \bibinfo {author} {\bibfnamefont {S.-J.}\ \bibnamefont {Chen}},
  \bibinfo {author} {\bibfnamefont {L.}~\bibnamefont {Zhang}}, \bibinfo
  {author} {\bibfnamefont {L.-X.}\ \bibnamefont {You}}, \bibinfo {author}
  {\bibfnamefont {Z.}~\bibnamefont {Wang}}, \bibinfo {author} {\bibfnamefont
  {Y.}~\bibnamefont {Liu}}, \bibinfo {author} {\bibfnamefont {C.-Y.}\
  \bibnamefont {Lu}}, \bibinfo {author} {\bibfnamefont {X.}~\bibnamefont
  {Jiang}}, \bibinfo {author} {\bibfnamefont {X.}~\bibnamefont {Ma}}, \bibinfo
  {author} {\bibfnamefont {Q.}~\bibnamefont {Zhang}}, \bibinfo {author}
  {\bibfnamefont {T.-Y.}\ \bibnamefont {Chen}}, \ and\ \bibinfo {author}
  {\bibfnamefont {J.-W.}\ \bibnamefont {Pan}},\ }\href@noop {} {\bibfield
  {journal} {\bibinfo  {journal} {Physical Review X}\ }\textbf {\bibinfo
  {volume} {6}},\ \bibinfo {pages} {011024} (\bibinfo {year}
  {2016}{\natexlab{a}})}\BibitemShut {NoStop}%
\bibitem [{\citenamefont {Xu}\ \emph {et~al.}(2014)\citenamefont {Xu},
  \citenamefont {Xu},\ and\ \citenamefont {Lo}}]{xu2014protocol}%
  \BibitemOpen
  \bibfield  {author} {\bibinfo {author} {\bibfnamefont {F.}~\bibnamefont
  {Xu}}, \bibinfo {author} {\bibfnamefont {H.}~\bibnamefont {Xu}}, \ and\
  \bibinfo {author} {\bibfnamefont {H.-K.}\ \bibnamefont {Lo}},\ }\href@noop {}
  {\bibfield  {journal} {\bibinfo  {journal} {Physical Review A}\ }\textbf
  {\bibinfo {volume} {89}},\ \bibinfo {pages} {052333} (\bibinfo {year}
  {2014})}\BibitemShut {NoStop}%
\bibitem [{\citenamefont {Curty}\ \emph {et~al.}(2014)\citenamefont {Curty},
  \citenamefont {Xu}, \citenamefont {Cui}, \citenamefont {Lim}, \citenamefont
  {Tamaki},\ and\ \citenamefont {Lo}}]{curty2014finite}%
  \BibitemOpen
  \bibfield  {author} {\bibinfo {author} {\bibfnamefont {M.}~\bibnamefont
  {Curty}}, \bibinfo {author} {\bibfnamefont {F.}~\bibnamefont {Xu}}, \bibinfo
  {author} {\bibfnamefont {W.}~\bibnamefont {Cui}}, \bibinfo {author}
  {\bibfnamefont {C.~C.~W.}\ \bibnamefont {Lim}}, \bibinfo {author}
  {\bibfnamefont {K.}~\bibnamefont {Tamaki}}, \ and\ \bibinfo {author}
  {\bibfnamefont {H.-K.}\ \bibnamefont {Lo}},\ }\href@noop {} {\bibfield
  {journal} {\bibinfo  {journal} {Nature Communications}\ }\textbf {\bibinfo
  {volume} {5}},\ \bibinfo {pages} {3732} (\bibinfo {year} {2014})}\BibitemShut
  {NoStop}%
\bibitem [{\citenamefont {Xu}\ \emph {et~al.}(2015)\citenamefont {Xu},
  \citenamefont {Curty}, \citenamefont {Qi}, \citenamefont {Qian},\ and\
  \citenamefont {Lo}}]{xu2015discrete}%
  \BibitemOpen
  \bibfield  {author} {\bibinfo {author} {\bibfnamefont {F.}~\bibnamefont
  {Xu}}, \bibinfo {author} {\bibfnamefont {M.}~\bibnamefont {Curty}}, \bibinfo
  {author} {\bibfnamefont {B.}~\bibnamefont {Qi}}, \bibinfo {author}
  {\bibfnamefont {L.}~\bibnamefont {Qian}}, \ and\ \bibinfo {author}
  {\bibfnamefont {H.-K.}\ \bibnamefont {Lo}},\ }\href@noop {} {\bibfield
  {journal} {\bibinfo  {journal} {Nature Photonics}\ }\textbf {\bibinfo
  {volume} {9}},\ \bibinfo {pages} {772} (\bibinfo {year} {2015})}\BibitemShut
  {NoStop}%
\bibitem [{\citenamefont {Pirandola}\ \emph {et~al.}(2015)\citenamefont
  {Pirandola}, \citenamefont {Ottaviani}, \citenamefont {Spedalieri},
  \citenamefont {Weedbrook}, \citenamefont {Braunstein}, \citenamefont {Lloyd},
  \citenamefont {Gehring}, \citenamefont {Jacobsen},\ and\ \citenamefont
  {Andersen}}]{CVDV}%
  \BibitemOpen
  \bibfield  {author} {\bibinfo {author} {\bibfnamefont {S.}~\bibnamefont
  {Pirandola}}, \bibinfo {author} {\bibfnamefont {C.}~\bibnamefont
  {Ottaviani}}, \bibinfo {author} {\bibfnamefont {G.}~\bibnamefont
  {Spedalieri}}, \bibinfo {author} {\bibfnamefont {C.}~\bibnamefont
  {Weedbrook}}, \bibinfo {author} {\bibfnamefont {S.~L.}\ \bibnamefont
  {Braunstein}}, \bibinfo {author} {\bibfnamefont {S.}~\bibnamefont {Lloyd}},
  \bibinfo {author} {\bibfnamefont {T.}~\bibnamefont {Gehring}}, \bibinfo
  {author} {\bibfnamefont {C.~S.}\ \bibnamefont {Jacobsen}}, \ and\ \bibinfo
  {author} {\bibfnamefont {U.~L.}\ \bibnamefont {Andersen}},\ }\href@noop {}
  {\bibfield  {journal} {\bibinfo  {journal} {Nature Photonics}\ }\textbf
  {\bibinfo {volume} {9}},\ \bibinfo {pages} {773} (\bibinfo {year}
  {2015})}\BibitemShut {NoStop}%
\bibitem [{\citenamefont {Zhou}\ \emph {et~al.}(2016)\citenamefont {Zhou},
  \citenamefont {Yu},\ and\ \citenamefont {Wang}}]{zhou2016making}%
  \BibitemOpen
  \bibfield  {author} {\bibinfo {author} {\bibfnamefont {Y.-H.}\ \bibnamefont
  {Zhou}}, \bibinfo {author} {\bibfnamefont {Z.-W.}\ \bibnamefont {Yu}}, \ and\
  \bibinfo {author} {\bibfnamefont {X.-B.}\ \bibnamefont {Wang}},\ }\href@noop
  {} {\bibfield  {journal} {\bibinfo  {journal} {Physical Review A}\ }\textbf
  {\bibinfo {volume} {93}},\ \bibinfo {pages} {042324} (\bibinfo {year}
  {2016})}\BibitemShut {NoStop}%
\bibitem [{\citenamefont {Rubenok}\ \emph {et~al.}(2013)\citenamefont
  {Rubenok}, \citenamefont {Slater}, \citenamefont {Chan}, \citenamefont
  {Lucio-Martinez},\ and\ \citenamefont {Tittel}}]{rubenok2013real}%
  \BibitemOpen
  \bibfield  {author} {\bibinfo {author} {\bibfnamefont {A.}~\bibnamefont
  {Rubenok}}, \bibinfo {author} {\bibfnamefont {J.~A.}\ \bibnamefont {Slater}},
  \bibinfo {author} {\bibfnamefont {P.}~\bibnamefont {Chan}}, \bibinfo {author}
  {\bibfnamefont {I.}~\bibnamefont {Lucio-Martinez}}, \ and\ \bibinfo {author}
  {\bibfnamefont {W.}~\bibnamefont {Tittel}},\ }\href@noop {} {\bibfield
  {journal} {\bibinfo  {journal} {Physical Review Letters}\ }\textbf {\bibinfo
  {volume} {111}},\ \bibinfo {pages} {130501} (\bibinfo {year}
  {2013})}\BibitemShut {NoStop}%
\bibitem [{\citenamefont {da~Silva}\ \emph {et~al.}(2013)\citenamefont
  {da~Silva}, \citenamefont {Vitoreti}, \citenamefont {Xavier}, \citenamefont
  {do~Amaral}, \citenamefont {Temporao},\ and\ \citenamefont {von~der
  Weid}}]{da2013proof}%
  \BibitemOpen
  \bibfield  {author} {\bibinfo {author} {\bibfnamefont {T.~F.}\ \bibnamefont
  {da~Silva}}, \bibinfo {author} {\bibfnamefont {D.}~\bibnamefont {Vitoreti}},
  \bibinfo {author} {\bibfnamefont {G.}~\bibnamefont {Xavier}}, \bibinfo
  {author} {\bibfnamefont {G.}~\bibnamefont {do~Amaral}}, \bibinfo {author}
  {\bibfnamefont {G.}~\bibnamefont {Temporao}}, \ and\ \bibinfo {author}
  {\bibfnamefont {J.}~\bibnamefont {von~der Weid}},\ }\href@noop {} {\bibfield
  {journal} {\bibinfo  {journal} {Physical Review A}\ }\textbf {\bibinfo
  {volume} {88}},\ \bibinfo {pages} {052303} (\bibinfo {year}
  {2013})}\BibitemShut {NoStop}%
\bibitem [{\citenamefont {Liu}\ \emph {et~al.}(2013)\citenamefont {Liu},
  \citenamefont {Chen}, \citenamefont {Wang}, \citenamefont {Liang},
  \citenamefont {Shentu}, \citenamefont {Wang}, \citenamefont {Cui},
  \citenamefont {Yin}, \citenamefont {Liu}, \citenamefont {Li}, \citenamefont
  {Ma}, \citenamefont {Pelc}, \citenamefont {Fejer}, \citenamefont {Peng},
  \citenamefont {Zhang},\ and\ \citenamefont {Pan}}]{liu2013experimental}%
  \BibitemOpen
  \bibfield  {author} {\bibinfo {author} {\bibfnamefont {Y.}~\bibnamefont
  {Liu}}, \bibinfo {author} {\bibfnamefont {T.-Y.}\ \bibnamefont {Chen}},
  \bibinfo {author} {\bibfnamefont {L.-J.}\ \bibnamefont {Wang}}, \bibinfo
  {author} {\bibfnamefont {H.}~\bibnamefont {Liang}}, \bibinfo {author}
  {\bibfnamefont {G.-L.}\ \bibnamefont {Shentu}}, \bibinfo {author}
  {\bibfnamefont {J.}~\bibnamefont {Wang}}, \bibinfo {author} {\bibfnamefont
  {K.}~\bibnamefont {Cui}}, \bibinfo {author} {\bibfnamefont {H.-L.}\
  \bibnamefont {Yin}}, \bibinfo {author} {\bibfnamefont {N.-L.}\ \bibnamefont
  {Liu}}, \bibinfo {author} {\bibfnamefont {L.}~\bibnamefont {Li}}, \bibinfo
  {author} {\bibfnamefont {X.}~\bibnamefont {Ma}}, \bibinfo {author}
  {\bibfnamefont {J.~S.}\ \bibnamefont {Pelc}}, \bibinfo {author}
  {\bibfnamefont {M.~M.}\ \bibnamefont {Fejer}}, \bibinfo {author}
  {\bibfnamefont {C.-Z.}\ \bibnamefont {Peng}}, \bibinfo {author}
  {\bibfnamefont {Q.}~\bibnamefont {Zhang}}, \ and\ \bibinfo {author}
  {\bibfnamefont {J.-W.}\ \bibnamefont {Pan}},\ }\href@noop {} {\bibfield
  {journal} {\bibinfo  {journal} {Physical Review Letters}\ }\textbf {\bibinfo
  {volume} {111}},\ \bibinfo {pages} {130502} (\bibinfo {year}
  {2013})}\BibitemShut {NoStop}%
\bibitem [{\citenamefont {Comandar}\ \emph {et~al.}(2016)\citenamefont
  {Comandar}, \citenamefont {Lucamarini}, \citenamefont {Fr{\"o}hlich},
  \citenamefont {Dynes}, \citenamefont {Sharpe}, \citenamefont {Tam},
  \citenamefont {Yuan}, \citenamefont {Penty},\ and\ \citenamefont
  {Shields}}]{comandar2016quantum}%
  \BibitemOpen
  \bibfield  {author} {\bibinfo {author} {\bibfnamefont {L.}~\bibnamefont
  {Comandar}}, \bibinfo {author} {\bibfnamefont {M.}~\bibnamefont
  {Lucamarini}}, \bibinfo {author} {\bibfnamefont {B.}~\bibnamefont
  {Fr{\"o}hlich}}, \bibinfo {author} {\bibfnamefont {J.}~\bibnamefont {Dynes}},
  \bibinfo {author} {\bibfnamefont {A.}~\bibnamefont {Sharpe}}, \bibinfo
  {author} {\bibfnamefont {S.-B.}\ \bibnamefont {Tam}}, \bibinfo {author}
  {\bibfnamefont {Z.}~\bibnamefont {Yuan}}, \bibinfo {author} {\bibfnamefont
  {R.~V.}\ \bibnamefont {Penty}}, \ and\ \bibinfo {author} {\bibfnamefont
  {A.}~\bibnamefont {Shields}},\ }\href@noop {} {\bibfield  {journal} {\bibinfo
   {journal} {Nature Photonics}\ }\textbf {\bibinfo {volume} {10}},\ \bibinfo
  {pages} {312} (\bibinfo {year} {2016})}\BibitemShut {NoStop}%
\bibitem [{\citenamefont {Tang}\ \emph
  {et~al.}(2016{\natexlab{b}})\citenamefont {Tang}, \citenamefont {Sun},
  \citenamefont {Xu}, \citenamefont {Chen}, \citenamefont {Li},\ and\
  \citenamefont {Liang}}]{tang2016experimental}%
  \BibitemOpen
  \bibfield  {author} {\bibinfo {author} {\bibfnamefont {G.-Z.}\ \bibnamefont
  {Tang}}, \bibinfo {author} {\bibfnamefont {S.-H.}\ \bibnamefont {Sun}},
  \bibinfo {author} {\bibfnamefont {F.}~\bibnamefont {Xu}}, \bibinfo {author}
  {\bibfnamefont {H.}~\bibnamefont {Chen}}, \bibinfo {author} {\bibfnamefont
  {C.-Y.}\ \bibnamefont {Li}}, \ and\ \bibinfo {author} {\bibfnamefont {L.-M.}\
  \bibnamefont {Liang}},\ }\href@noop {} {\bibfield  {journal} {\bibinfo
  {journal} {Physical Review A}\ }\textbf {\bibinfo {volume} {94}},\ \bibinfo
  {pages} {032326} (\bibinfo {year} {2016}{\natexlab{b}})}\BibitemShut
  {NoStop}%
\bibitem [{\citenamefont {Jain}\ \emph
  {et~al.}(2011{\natexlab{b}})\citenamefont {Jain}, \citenamefont {Wittmann},
  \citenamefont {Lydersen}, \citenamefont {Wiechers}, \citenamefont {Elser},
  \citenamefont {Marquardt}, \citenamefont {Makarov},\ and\ \citenamefont
  {Leuchs}}]{Jain2011Device}%
  \BibitemOpen
  \bibfield  {author} {\bibinfo {author} {\bibfnamefont {N.}~\bibnamefont
  {Jain}}, \bibinfo {author} {\bibfnamefont {C.}~\bibnamefont {Wittmann}},
  \bibinfo {author} {\bibfnamefont {L.}~\bibnamefont {Lydersen}}, \bibinfo
  {author} {\bibfnamefont {C.}~\bibnamefont {Wiechers}}, \bibinfo {author}
  {\bibfnamefont {D.}~\bibnamefont {Elser}}, \bibinfo {author} {\bibfnamefont
  {C.}~\bibnamefont {Marquardt}}, \bibinfo {author} {\bibfnamefont
  {V.}~\bibnamefont {Makarov}}, \ and\ \bibinfo {author} {\bibfnamefont
  {G.}~\bibnamefont {Leuchs}},\ }\href@noop {} {\bibfield  {journal} {\bibinfo
  {journal} {Physical Review Letters}\ }\textbf {\bibinfo {volume} {107}},\
  \bibinfo {pages} {110501} (\bibinfo {year} {2011}{\natexlab{b}})}\BibitemShut
  {NoStop}%
\bibitem [{\citenamefont {Yin}\ \emph {et~al.}(2014)\citenamefont {Yin},
  \citenamefont {Wang}, \citenamefont {Chen}, \citenamefont {Li}, \citenamefont
  {Guo},\ and\ \citenamefont {Han}}]{yin2014reference}%
  \BibitemOpen
  \bibfield  {author} {\bibinfo {author} {\bibfnamefont {Z.-Q.}\ \bibnamefont
  {Yin}}, \bibinfo {author} {\bibfnamefont {S.}~\bibnamefont {Wang}}, \bibinfo
  {author} {\bibfnamefont {W.}~\bibnamefont {Chen}}, \bibinfo {author}
  {\bibfnamefont {H.-W.}\ \bibnamefont {Li}}, \bibinfo {author} {\bibfnamefont
  {G.-C.}\ \bibnamefont {Guo}}, \ and\ \bibinfo {author} {\bibfnamefont
  {Z.-F.}\ \bibnamefont {Han}},\ }\href@noop {} {\bibfield  {journal} {\bibinfo
   {journal} {Quantum Information Processing}\ }\textbf {\bibinfo {volume}
  {13}},\ \bibinfo {pages} {1237} (\bibinfo {year} {2014})}\BibitemShut
  {NoStop}%
\bibitem [{\citenamefont {Wang}\ \emph {et~al.}(2015)\citenamefont {Wang},
  \citenamefont {Song}, \citenamefont {Yin}, \citenamefont {Wang},
  \citenamefont {Chen}, \citenamefont {Zhang}, \citenamefont {Guo},\ and\
  \citenamefont {Han}}]{wang2015phase}%
  \BibitemOpen
  \bibfield  {author} {\bibinfo {author} {\bibfnamefont {C.}~\bibnamefont
  {Wang}}, \bibinfo {author} {\bibfnamefont {X.-T.}\ \bibnamefont {Song}},
  \bibinfo {author} {\bibfnamefont {Z.-Q.}\ \bibnamefont {Yin}}, \bibinfo
  {author} {\bibfnamefont {S.}~\bibnamefont {Wang}}, \bibinfo {author}
  {\bibfnamefont {W.}~\bibnamefont {Chen}}, \bibinfo {author} {\bibfnamefont
  {C.-M.}\ \bibnamefont {Zhang}}, \bibinfo {author} {\bibfnamefont {G.-C.}\
  \bibnamefont {Guo}}, \ and\ \bibinfo {author} {\bibfnamefont {Z.-F.}\
  \bibnamefont {Han}},\ }\href@noop {} {\bibfield  {journal} {\bibinfo
  {journal} {Physical Review Letters}\ }\textbf {\bibinfo {volume} {115}},\
  \bibinfo {pages} {160502} (\bibinfo {year} {2015})}\BibitemShut {NoStop}%
\bibitem [{\citenamefont {Wang}\ \emph
  {et~al.}(2017{\natexlab{a}})\citenamefont {Wang}, \citenamefont {Yin},
  \citenamefont {Wang}, \citenamefont {Chen}, \citenamefont {Guo},\ and\
  \citenamefont {Han}}]{wang2017measurement}%
  \BibitemOpen
  \bibfield  {author} {\bibinfo {author} {\bibfnamefont {C.}~\bibnamefont
  {Wang}}, \bibinfo {author} {\bibfnamefont {Z.-Q.}\ \bibnamefont {Yin}},
  \bibinfo {author} {\bibfnamefont {S.}~\bibnamefont {Wang}}, \bibinfo {author}
  {\bibfnamefont {W.}~\bibnamefont {Chen}}, \bibinfo {author} {\bibfnamefont
  {G.-C.}\ \bibnamefont {Guo}}, \ and\ \bibinfo {author} {\bibfnamefont
  {Z.-F.}\ \bibnamefont {Han}},\ }\href@noop {} {\bibfield  {journal} {\bibinfo
   {journal} {Optica}\ }\textbf {\bibinfo {volume} {4}},\ \bibinfo {pages}
  {1016} (\bibinfo {year} {2017}{\natexlab{a}})}\BibitemShut {NoStop}%
\bibitem [{\citenamefont {Zhang}\ \emph
  {et~al.}(2017{\natexlab{a}})\citenamefont {Zhang}, \citenamefont {Zhu},\ and\
  \citenamefont {Wang}}]{zhang2017practical}%
  \BibitemOpen
  \bibfield  {author} {\bibinfo {author} {\bibfnamefont {C.-M.}\ \bibnamefont
  {Zhang}}, \bibinfo {author} {\bibfnamefont {J.-R.}\ \bibnamefont {Zhu}}, \
  and\ \bibinfo {author} {\bibfnamefont {Q.}~\bibnamefont {Wang}},\ }\href@noop
  {} {\bibfield  {journal} {\bibinfo  {journal} {Physical Review A}\ }\textbf
  {\bibinfo {volume} {95}},\ \bibinfo {pages} {032309} (\bibinfo {year}
  {2017}{\natexlab{a}})}\BibitemShut {NoStop}%
\bibitem [{\citenamefont {Zhang}\ \emph
  {et~al.}(2017{\natexlab{b}})\citenamefont {Zhang}, \citenamefont {Zhu},\ and\
  \citenamefont {Wang}}]{zhang2017decoy}%
  \BibitemOpen
  \bibfield  {author} {\bibinfo {author} {\bibfnamefont {C.-M.}\ \bibnamefont
  {Zhang}}, \bibinfo {author} {\bibfnamefont {J.-R.}\ \bibnamefont {Zhu}}, \
  and\ \bibinfo {author} {\bibfnamefont {Q.}~\bibnamefont {Wang}},\ }\href@noop
  {} {\bibfield  {journal} {\bibinfo  {journal} {Journal of Lightwave
  Technology}\ }\textbf {\bibinfo {volume} {35}},\ \bibinfo {pages} {4574}
  (\bibinfo {year} {2017}{\natexlab{b}})}\BibitemShut {NoStop}%
\bibitem [{\citenamefont {Tang}\ \emph
  {et~al.}(2016{\natexlab{c}})\citenamefont {Tang}, \citenamefont {Sun},
  \citenamefont {Chen}, \citenamefont {Li},\ and\ \citenamefont
  {Liang}}]{tang2016time}%
  \BibitemOpen
  \bibfield  {author} {\bibinfo {author} {\bibfnamefont {G.-Z.}\ \bibnamefont
  {Tang}}, \bibinfo {author} {\bibfnamefont {S.-H.}\ \bibnamefont {Sun}},
  \bibinfo {author} {\bibfnamefont {H.}~\bibnamefont {Chen}}, \bibinfo {author}
  {\bibfnamefont {C.-Y.}\ \bibnamefont {Li}}, \ and\ \bibinfo {author}
  {\bibfnamefont {L.-M.}\ \bibnamefont {Liang}},\ }\href@noop {} {\bibfield
  {journal} {\bibinfo  {journal} {Chinese Physics Letters}\ }\textbf {\bibinfo
  {volume} {33}},\ \bibinfo {pages} {120301} (\bibinfo {year}
  {2016}{\natexlab{c}})}\BibitemShut {NoStop}%
\bibitem [{\citenamefont {Tang}\ \emph {et~al.}(2013)\citenamefont {Tang},
  \citenamefont {Yin}, \citenamefont {Ma}, \citenamefont {Fung}, \citenamefont
  {Liu}, \citenamefont {Yong}, \citenamefont {Chen}, \citenamefont {Peng},
  \citenamefont {Chen},\ and\ \citenamefont {Pan}}]{tang2013source}%
  \BibitemOpen
  \bibfield  {author} {\bibinfo {author} {\bibfnamefont {Y.-L.}\ \bibnamefont
  {Tang}}, \bibinfo {author} {\bibfnamefont {H.-L.}\ \bibnamefont {Yin}},
  \bibinfo {author} {\bibfnamefont {X.}~\bibnamefont {Ma}}, \bibinfo {author}
  {\bibfnamefont {C.-H.~F.}\ \bibnamefont {Fung}}, \bibinfo {author}
  {\bibfnamefont {Y.}~\bibnamefont {Liu}}, \bibinfo {author} {\bibfnamefont
  {H.-L.}\ \bibnamefont {Yong}}, \bibinfo {author} {\bibfnamefont {T.-Y.}\
  \bibnamefont {Chen}}, \bibinfo {author} {\bibfnamefont {C.-Z.}\ \bibnamefont
  {Peng}}, \bibinfo {author} {\bibfnamefont {Z.-B.}\ \bibnamefont {Chen}}, \
  and\ \bibinfo {author} {\bibfnamefont {J.-W.}\ \bibnamefont {Pan}},\
  }\href@noop {} {\bibfield  {journal} {\bibinfo  {journal} {Physical Review
  A}\ }\textbf {\bibinfo {volume} {88}},\ \bibinfo {pages} {022308} (\bibinfo
  {year} {2013})}\BibitemShut {NoStop}%
\bibitem [{\citenamefont {Wang}\ \emph
  {et~al.}(2017{\natexlab{b}})\citenamefont {Wang}, \citenamefont {Wang},
  \citenamefont {Chen}, \citenamefont {Wang}, \citenamefont {Chen},
  \citenamefont {Yin}, \citenamefont {He}, \citenamefont {Guo},\ and\
  \citenamefont {Han}}]{Wang:17}%
  \BibitemOpen
  \bibfield  {author} {\bibinfo {author} {\bibfnamefont {C.}~\bibnamefont
  {Wang}}, \bibinfo {author} {\bibfnamefont {F.-X.}\ \bibnamefont {Wang}},
  \bibinfo {author} {\bibfnamefont {H.}~\bibnamefont {Chen}}, \bibinfo {author}
  {\bibfnamefont {S.}~\bibnamefont {Wang}}, \bibinfo {author} {\bibfnamefont
  {W.}~\bibnamefont {Chen}}, \bibinfo {author} {\bibfnamefont {Z.-Q.}\
  \bibnamefont {Yin}}, \bibinfo {author} {\bibfnamefont {D.-Y.}\ \bibnamefont
  {He}}, \bibinfo {author} {\bibfnamefont {G.-C.}\ \bibnamefont {Guo}}, \ and\
  \bibinfo {author} {\bibfnamefont {Z.-F.}\ \bibnamefont {Han}},\ }\href
  {http://jlt.osa.org/abstract.cfm?URI=jlt-35-23-4996} {\bibfield  {journal}
  {\bibinfo  {journal} {J. Lightwave Technol.}\ }\textbf {\bibinfo {volume}
  {35}},\ \bibinfo {pages} {4996} (\bibinfo {year}
  {2017}{\natexlab{b}})}\BibitemShut {NoStop}%
\bibitem [{\citenamefont {Ma}\ and\ \citenamefont
  {Razavi}(2012)}]{Ma2012Alternative}%
  \BibitemOpen
  \bibfield  {author} {\bibinfo {author} {\bibfnamefont {X.}~\bibnamefont
  {Ma}}\ and\ \bibinfo {author} {\bibfnamefont {M.}~\bibnamefont {Razavi}},\
  }\href@noop {} {\bibfield  {journal} {\bibinfo  {journal} {Physical Review
  A}\ }\textbf {\bibinfo {volume} {86}},\ \bibinfo {pages} {3818} (\bibinfo
  {year} {2012})}\BibitemShut {NoStop}%
\bibitem [{\citenamefont {Yu}\ \emph {et~al.}(2013)\citenamefont {Yu},
  \citenamefont {Zhou},\ and\ \citenamefont {Wang}}]{yu2013three}%
  \BibitemOpen
  \bibfield  {author} {\bibinfo {author} {\bibfnamefont {Z.-W.}\ \bibnamefont
  {Yu}}, \bibinfo {author} {\bibfnamefont {Y.-H.}\ \bibnamefont {Zhou}}, \ and\
  \bibinfo {author} {\bibfnamefont {X.-B.}\ \bibnamefont {Wang}},\ }\href@noop
  {} {\bibfield  {journal} {\bibinfo  {journal} {Physical Review A}\ }\textbf
  {\bibinfo {volume} {88}},\ \bibinfo {pages} {062339} (\bibinfo {year}
  {2013})}\BibitemShut {NoStop}%
\end{thebibliography}%

\end{document}